\documentclass[times,doublespace]{fldauth}

\usepackage{moreverb}

\usepackage{amssymb,bm,amsmath}

\usepackage[numbers,sort]{natbib}

\usepackage[active]{correct}

\newcommand\BibTeX{{\rmfamily B\kern-.05em \textsc{i\kern-.025em b}\kern-.08em
T\kern-.1667em\lower.7ex\hbox{E}\kern-.125emX}}

\def\volumeyear{20xx}

\begin{document}

\runningheads{J.~M.~Alam {\em et al}}{A simple and efficient discretization  using a dyadic refinement approach}

\title{A computational methodology for two-dimensional fluid flows}

\author{Jahrul~M~Alam\corrauth, Raymond P Walsh, M Alamgir Hossain, Andrew M Rose}

\address{Department of Mathematics and Statistics, Memorial University, Canada, A1C 5S7}

\corraddr{Department of Mathematics and Statistics, Memorial University, Canada, A1C 5S7}

\begin{abstract}
A weighted residual collocation methodology for simulating two-dimensional shear driven and natural convection flows has been presented. Using a dyadic mesh refinement, the methodology generates a basis and a multiresolution scheme to approximate a fluid flow. To extend the benefits of the dyadic mesh refinement approach to the field of computational fluid dynamics, this article has studied an iterative interpolation scheme for the construction and differentiation of a basis function in a two-dimensional mesh that is a finite collection of rectangular elements. We have verified that, on a given mesh, the discretization error is controlled by the order of the basis function. The potential of this novel technique has been demonstrated with some representative examples of the Poisson equation. We have also verified the technique with a dynamical core of a two-dimensional flow in primitive variables. An excellent result has been observed -- on resolving a shear layer and on the conservation of the potential and the kinetic energies -- with respect to previously reported benchmark simulations. In particular, the shear driven simulation at $\hbox{CFL}=2.5$ (Courant~Friedrichs~Lewy) and $\mathcal Re=1\,000$ (Reynolds number) exhibits a linear speed up of CPU time with an increase of the time step, $\Delta t$. For the natural convection flow, the conversion of the potential energy to the kinetic energy and the conservation of total energy is resolved by the proposed method. The computed streamlines and the velocity fields have been demonstrated. \\ 
\end{abstract}


\maketitle


\section{Introduction}
The overall quality of a computational fluid dynamics~(CFD) simulation is influenced by the appropriate discrete representation of the continuum mechanics (see, part I of~\cite{Tannehill97}), as well as by the solution of the Poisson equation for the pressure~\cite{Pozrikidis2001,Gresho87,San2013,Weinan95,Jobelin2006,Laizet2009}. An optimal discretization -- that resolves nonlinear advection and diffusion of momentum as well as the dependence between the velocity and the pressure -- remains challenging~\cite{Wesseling2000}. 
Obtaining an efficient and accurate solution of the Poisson equation~\cite{Wesseling2004} and the Navier-Stokes equation~\cite{Wesseling2000,Tannehill97} is a longstanding challenge, and is an active interdisciplinary research topic. For example, in chemical engineering, the Poisson equation models the electrostatic potential of an electric field with continuously distributed charges~\cite{Patankar98,Alam2002,Alam2012b,Fogo2002,Jackson3rd,Roux91,Genovese2007}.  A solution is often obtained by evaluating the integral of the charge distribution using the fast multiple method~\cite{Greeng97}, and this technique is also used by the vortex method algorithm~\cite{Petros2000,Petros2005}. In Fluid Dynamics, a Poisson equation is solved for computing the divergence free velocity in pressure-based approaches (pressure-Poisson equation). A `Poisson like' nonlinear vector equation  (or nonlinear Poisson equation, e.g. \cite{Dorfler95}) may also be obtained from the implicit in time discretization of the momentum equation~\cite{Alam2011,Alam2012,Choi94,Tannehill97,Tuckerman89}. In the algorithm presented by~\citet{Choi94}, a `Poisson like' nonlinear vector equation was solved for the velocity, and a scalar Poisson equation was solved for the pressure. In a velocity-pressure approach, although iterative techniques -- such as the multigrid method~\cite{Wesseling2004} -- provides a rapid computing algorithm, in heat and mass transfer analysis~(e.g.~\cite{Alam2012}), the scalar Poisson equation takes more computational overhead than the accompanying vector advection-diffusion equation of the system (see~\cite{San2013} for a comprehensive review). \Add{Many other authors show a growing interest on improving computational efficiency for similar problems~\cite{Jobelin2006,Perrin2006,Kannan2009,Kai2010,Kannan2010,Kannan2011,Zeng2012}.}

In the present approach, we discretize \Add{the set of nonlinear partial differential equations, governing the conservation of mass, momentum, and enrgy} of a fluid flow with the implicit Crank-Nicolson method, which leads to a nonlinear system of `Poisson like' equations. \Add{For this implicit treatment, there is no restriction on the time step to ensure the stability of the time integration scheme~(see, \cite{Choi94,Tannehill97}).} The pressure is diagnosed from the density {\em via} the equation of state (see, \citet{Perrin2006}). For example, in a two-dimensional shear driven flow, the velocities ($u$, $v$) and the density ($\rho$) are computed iteratively, where the variables are nonlinearly coupled. In other words, the physics of the flow is approximated at each iteration until a convergence is reached. 
Considering benchmark CFD examples, we have presented verification of the methodology in terms of implementation easiness, computational efficiency, and improvement in the modelled physics. Note that we do not aim to address the challenges associated with the classical pressure Poisson equation - as described by~\citet{San2013} and \citet{Pozrikidis2001}; however, we want to study an alternative algorithm that applies a Poisson solver to a `Poisson like' nonlinear vector equation.

This article thus focuses on the study of a weighted residual collocation method to approximate partial derivatives of the Poisson equation on nested multiresolution meshes, which is based on the construction of a smooth surface using an iterative interpolation. This iterative interpolation is also known as a subdivision scheme (see, \cite{Sweldens97,Shen97}). It is built on the fundamental function and multiresolution refinement technique of~\citet{Dubuc89} ({\em hereinafter}, DD subdivision). Although the DD subdivision was shown powerful and efficient in many applications, its simple genesis is often deceptive. More importantly, DD subdivision was not exploited fully to approximate derivatives, even though it was used to enhance techniques of solving partial differential equations~(PDEs) ({\em e.g.}~\cite{Oleg2005,Mehra2008}). 

The beautiful subdivision process is also a primary mechanism for the second generation wavelets, as well as for the multiresolution best $\mathcal N$ term approximation schemes~\cite{DeVore98,Sweldens95}. In the past decade, the adaptive wavelet collocation method~(AWCM) (see, \cite{Oleg2005,Kai2010}) adopted the DD subdivision to construct a wavelet basis in order to achieve fine and coarse resolutions, locally and dynamically, where it is necessary~\cite{Alam2011,Alam2006,Oleg2000,Oleg2005}. However, in AWCM, partial derivatives are computed with a differential quadrature or a classical finite difference method~\cite{Oleg95,Oleg96,Oleg97,Jameson98}. A locally refined resolution may also be achieved without the subdivision scheme; {\em e.g.} see the recent multiresolution ocean model of~\citet{Ringler2013}. The pioneering multiresolution technique of~\citet{Brandt77} was verified efficient on many CFD simulations~\cite{Wesseling93,Wesseling2000,Wesseling2004}. In general, most of these multiresolution (or multilevel) approaches focused on robust techniques of solving the discretized system, where an existing discretization was used. In contrast, the present article demonstrates a multiresolution collocation approach to approximate derivatives, where the efficiency of solving the discrete system is obtained by taking benefits of some existing powerful Krylov space techniques~\cite{Brown94,Wesseling2000}. A brief literature review indicates that the discretization technique presented in this article has not been studied for solving the Poisson equation or the Navier-Stokes equation, although methods similar to the present approach were investigated~\cite{Oleg2005,Genovese2007,Harten97}. The novelty of the present work may also be hidden behind the simplicity of the DD scheme; however, the application to complex geometry/irregular mesh is one important step forward of this development with respect to equivalent, commonly used finite difference methods or rarely used differential quadrature methods -- although the present article has not included a rigorous study on complex geometry/irregular mesh.  

Generally speaking, the robustness of a computational technique depends on the accurate discretization and on a rapid algorithm for solving the discrete system. With a given number of nodes ($\mathcal N$), the present method reaches a higher order accuracy by increasing the order of DD scheme with $\mathcal O(\mathcal N)$ complexity (see, Table~\ref{tab:ex1ep}). However, in a fluid flow simulation, an appropriate measure of robustness may be how accurately the method resolves some conserved quantities or a flow specific property, such as a shear layer, and how fast the global solution is obtained ({\em e.g.} how large a time step, $\Delta t$, is). In Fluid Dynamics, \Add{the overall quality of a numerical simulation may also be affected by how the technique approaches to resolve the dependence between velocity and pressure, and readers may find a comprehensive discussion on both the pressure-based and density-based approaches in Chapter~9 of~\citet{Tannehill97}. We have considered these points to verify the present development using a density based approach.} Our numerical experiments indicate that, for a natural convection flow, conversion of the potential energy to the kinetic energy is captured sufficiently, \Add{without using a upwind/downwind stencil for the discretization of nonlinear advection terms or without implementing any artificial damping mechanism} (see, \cite{Tannehill97}). This is an important achievement of the present simulation approach. 

A weighted residual collocation method for the discretization of PDEs may be developed using an appropriate scaling function~\cite{Bruce,Mallat2009}. However, existing literature indicates that such a development remains open, although the scaling function was used by many authors. For example, \citet{Genovese2007} approximated the charge distribution (see, eq~(\ref{eq:bvp})) using the scaling function, and solved the Poisson equation for the electrostatic potential by evaluating an integral of the Green's function. \citet{Oleg2005} applied a differential quadrature on a multi-level grid for solving two- and three-dimensional Poisson equations on rectangles and rectangular prisms, using a wavelet basis, which is generated from the scaling function. \citet{Mehra2008} extended this multi-level differential quadrature AWCM for solving PDEs on a sphere, where the scaling function is extended to build spherical wavelets. Clearly, the DD subdivision was not fully explored to approximate partial derivatives in a more general domain.

The present weighted residual collocation method has been implemented in a domain that is a finite collection of rectangles or rectangular prisms. Such a domain needs  boundaries parallel to coordinate axes, and may contain holes. First, we study how to approximate the Laplacian
\begin{equation}
  \label{eq:bvp}
  \nabla^2P=\rho\quad\hbox{in}\quad\Omega\subseteq\mathbb R^d,\,(d=1,\,2,\,3),
\end{equation}
on a finite collection of $\mathcal N$ nodes using the interpolating scaling function, such that the potential $P(\bm x)$, $\bm x\in\Omega$, is represented by a smooth surface. For $\mathcal N\rightarrow\infty$, a sequence of such surfaces converges to the actual surface of the potential $P(x,y)$. Second, we study the solution of~(\ref{eq:bvp}) (along with appropriate boundary conditions) and some of its applications, where $\rho(\bm x)$ is given, and $P(\bm x)$ is desired. Third, we extend the methodology to discretize `Poisson like' nonlinear vector equations (where $\rho(P)$), {\em i.e.} the time discretized Navier-Stokes equation, and to simulate shear driven and natural convection circulations.

In section~\ref{sec:cmi}, the weighted residual collocation method and the interpolating scaling function is studied briefly. However, without presenting the abstract theoretical material (see, \cite{Mallat2009}), we have outlined the methodology for readers who may be interested in similar applications. Section~\ref{sec:pde} extends the developed methodology to solve PDEs, where we verify that the methodology resolves the energy conversion cycle between potential and kinetic energies. The present research has been summarized in section~\ref{sec:conc}.

\section{The collocation method and the interpolating scaling function}
\label{sec:cmi}
The present multiresolution collocation method is based on a set $\{\varphi_k(\bm x)\}$ of two-dimensional scaling functions (see, pp~$267$,~\cite{Mallat2009}). The basis is obtained by extending the one-dimensional fundamental function, $\varphi(x)$, which is uniformly continuous and at most twice differentiable~\cite{Dubuc89}. The extension is achieved by applying the DD subdivision on two-dimensional meshes~\cite{Dubuc91}. The partial derivatives are approximated by considering a trial solution that is spanned by the set $\{\varphi_k(\bm x)\}$~\cite{Bruce}. 

\subsection{The multiresolution collocation method}
Collocation methods are special cases of weighted residual methods~\cite{Bruce}. 
Consider two sets $\{\varphi_k(\bm x)\}$ and $\{\tilde\varphi_k(\bm x)\}$ of functions. To approximate the Laplacian~(\ref{eq:bvp}), the method of weighted residual considers the $\mathcal N$ term trial solution
\begin{equation}
  \label{eq:trl}
  P^{\mathcal N}(\bm x) = \sum_{k=0}^{\mathcal N-1}c_k\varphi_k(\bm x)
\end{equation}
over the basis $\{\varphi_k(\bm x)\}$, and assumes that the inner product
$$
\left\langle r(\bm x),\tilde\varphi_k(\bm x)\right\rangle =
\int\limits_{\Omega} r(\bm x)\tilde\varphi_k(\bm x)d\bm x
$$
vanishes, where 
$
  r(\bm x) = \rho(\bm x) - \nabla^2P^{\mathcal N}(\bm x)
$
is the residual and $\tilde\varphi_k(\bm x)$ are test functions. The functions $\varphi_k(\bm x)$ are chosen so that the trial solution~(\ref{eq:trl}) represents the $\mathcal N$ term best approximation of $P(\bm x)$. As described by~\citet{Bruce}, if the test functions $\tilde\varphi_k(\bm x)$ are the same as the trial functions $\varphi_k(\bm x)$, the weighted residual method takes the form of the best approximation method developed by~\citet{Galerkin15}. 
In contrast, the collocation method takes $\tilde\varphi_k(\bm x) = \delta(\bm x-\bm x_k)$ with respect to a set of nodes $\{\bm x_k\}$ in a domain, $\Omega$ (see,~\cite{Bruce}). 
As a result, 
$
\left\langle r(\bm x),\tilde\varphi_k(\bm x)\right\rangle =0
$
implies that the error of approximating $\nabla^2P(\bm x)$ by the trial solution $P^{\mathcal N}(\bm x)$ is exactly zero on all selected nodes $\bm x_k$~\cite{Bruce}. In a multiresolution approach~\cite{Mallat2009}, new nodes may be inserted dyadically into the old mesh (see, Fig~\ref{fig:ngh}$(a)$), and thus, the residual approaches zero everywhere when $\mathcal N\rightarrow\infty$ by the mesh refinement.

Collocation methods vary with the choice of the functions $\varphi_k(\bm x)$ (see, \cite{Bruce,Roger2002}). In this work, we study custom designed interpolating scaling functions, $\varphi_k(\bm x_j)=\delta_{kj}$, to develop a collocation method for PDEs on rectangular meshes. Since the residual, $r(\bm x)$, vanishes on each node, we get 
%
\begin{equation}
  \label{eq:sys}
\sum_{k}^{}\nabla^2\varphi_k(\bm x_j)c_k = \rho(\bm x_j),\quad j=0\ldots\mathcal N-1,
\end{equation}
where $k$ and $j$ are the indices of corresponding nodes. If the scaling functions $\varphi_k(\bm x)$ are exactly known or their derivatives are given, then the system~(\ref{eq:sys}) can be inverted -- along with suitable boundary conditions -- to find $c_k$'s, and an approximate solution of~(\ref{eq:bvp}) is given by the trial solution~(\ref{eq:trl}). 
The multiresolution collocation approximation is the following. 

The trial solution~(\ref{eq:trl}) is chosen from any of the nested approximation spaces
$$\mathcal V^0\subseteq\ldots\subseteq\mathcal V^{s-1}\subseteq\mathcal V^s\subseteq\mathcal V^{s+1}\ldots
\quad\hbox{and}\quad
\cup_{s=0}^{\infty}\mathcal V^s = L_2(\overline\Omega),
$$
where each $\mathcal V^s$ has a Riesz basis $\{\varphi_k(\bm x)\}$ and contains functions which may not oscillate at a frequency larger than $2^{s-1}$. The collection of approximation spaces $\{\mathcal V^s\}$ is called a multiresolution approximation space~\cite{Sweldens97}. A dual multiresolution approximation is a collection of spaces $\{\tilde{\mathcal V}^s\}$ with a Riesz basis $\{\tilde\varphi_k(\bm x)\}$, where $\langle\varphi_k(\bm x),\tilde\varphi_j(\bm x)\rangle = \delta_{k,j}$. A full theoretical details of the multiresolution approximation is given by~\citet{Mallat2009} (see, \cite{Sweldens97} and section~$7.1.1$ of~\cite{Mallat2009}). The trial function $P^{\mathcal N}(\bm x)$ defined by~(\ref{eq:trl}) is called a multiresolution projection of $P(\bm x)$ onto the space $\mathcal V^s$, where $\mathcal N$ is the dimension of $\mathcal V^s$ and $c_k=\langle P(\bm x),\tilde\varphi(\bm x)\rangle$. The trial function~(\ref{eq:trl}) has two possible representations~\cite{Mallat2009},
$$
P^{\mathcal N}(\bm x) = \sum\limits_{k=0}\limits^{\mathcal N-1}
\overbrace{
\left\langle P(\bm x),\tilde\varphi_k(\bm x)\right\rangle
}^{c_k}
\varphi_k(\bm x) =
\sum\limits_{k=0}\limits^{\mathcal N-1} 
\left\langle P(\bm x),\varphi_k(\bm x)\right\rangle
\tilde\varphi_k(\bm x), 
$$
satisfying the Riesz stability criterion.
Note the choice of $\tilde\varphi(\bm x) = \delta(\bm x-\bm x_k)$ in the present collocation method. Thus, eq~(\ref{eq:trl}) and the associated multiresoluion approximation live on a solid mathematical foundation, where the present article deals with its application.

The self-similarity of $\{\mathcal V^s\}$ in position suggests that -- in a collocation method -- these function spaces can be associated with a nested sequence of $b$-adic meshes (see, section~$7.8$ of~\cite{Mallat2009}),
$$\mathcal G^0\subseteq\ldots\subseteq\mathcal G^{s-1}\subseteq\mathcal G^s\subseteq\mathcal G^{s+1}\ldots
\quad\hbox{and}\quad
\lim\limits_{s\rightarrow\infty}\mathcal G^s = \overline\Omega.
$$
Such a mesh, $\mathcal G^s$, is a finite collection of elements -- rectangles in the present article -- and has a total of $\mathcal N$ nodes $\bm x_k$ for $k=0,\ldots,\mathcal N-1$ such that $\bm x_k\in\Omega\subseteq\mathbb R^d$ (for rectangles, $d=2$). The mesh $\mathcal G^{s+1}$ is obtained by refining elements of the mesh $\mathcal G^s$ with a factor of $b$ in each direction. 
In a $b$-adic refinement, an element gets $b^d$ child elements, and the mesh refinement can be managed efficiently with a tree data structure or by an existing mesh generation library. In this article, we have considered two-dimensional ($d=2$) meshes with $b=2$ unless otherwise stated. \Add{Let $\mathcal G^0$ be a 2D mesh of $m_x\times m_y$ nodes. For any fixed integer $s>0$, we get a mesh $\mathcal G^s$ with $\mathcal N = [(m_x-1)2^s+1]\times [(m_y-1)2^s+1]$ nodes. In such a mesh,} each node $\bm x_k$ may also be represented by $\bm x_{ij}$ for $0\le i\le n_x$ and $0\le j\le n_y$.
With respect to the index $k$ of the mesh $\mathcal G^s$, one notes that $\bm x_k\in\mathcal G^s$ and $\bm x_{2k}\in\mathcal G^{s+1}$ represent the same node because the meshes are nested, {\em i.e.} $\mathcal G^s\subseteq\mathcal G^{s+1}$. In other words, each node $\bm x_{2k}\in\mathcal G^{s+1}$ is present in the coarser mesh, and has $2^d-1$ neighbors $\bm x_{2k+1}\in\mathcal G^{s+1}$ those were not present in the coarser mesh. For example, neighbors of $\bm x_{2k} = \bm x_{2i,2j}$ are $\bm x_{2k+1} = \{\bm x_{2i+1,2j}, \bm x_{2i,2j+1}, \bm x_{2i+1,2j+1}\}$ with some exceptions on the boundaries.

In the present development, an iterative interpolation process, which is fully understood as a powerful numerical tool (see, \cite{Dubuc89}), has been employed for constructing a set $\{\varphi_k(\bm x)\}$ of interpolating scaling functions as the basis of the approximation space $\mathcal V^s$.

\subsection{DD interpolation of order $p$}

A detailed theory of the simple and powerful iterative interpolation and multiresolution refinement techniques was studied by many authors~\cite{Dubuc91,Dubuc89,Donoho96,Dubuc95,Shen97}. \citet{Dubuc89} developed the fundamental function through the iterative interpolation based on $2p$ collocation points. Readers may find further details of the continuity and the differentiablity of the fundamental function from works of~\citet{Dubuc89} and~\citet{Dubuc95}. We have outlined the process briefly based on a two-dimensional mesh; however, we have included both one- and two-dimensional examples.

\begin{figure}
  \begin{center}
  \begin{tabular}{cc}
    \includegraphics[height=4cm]{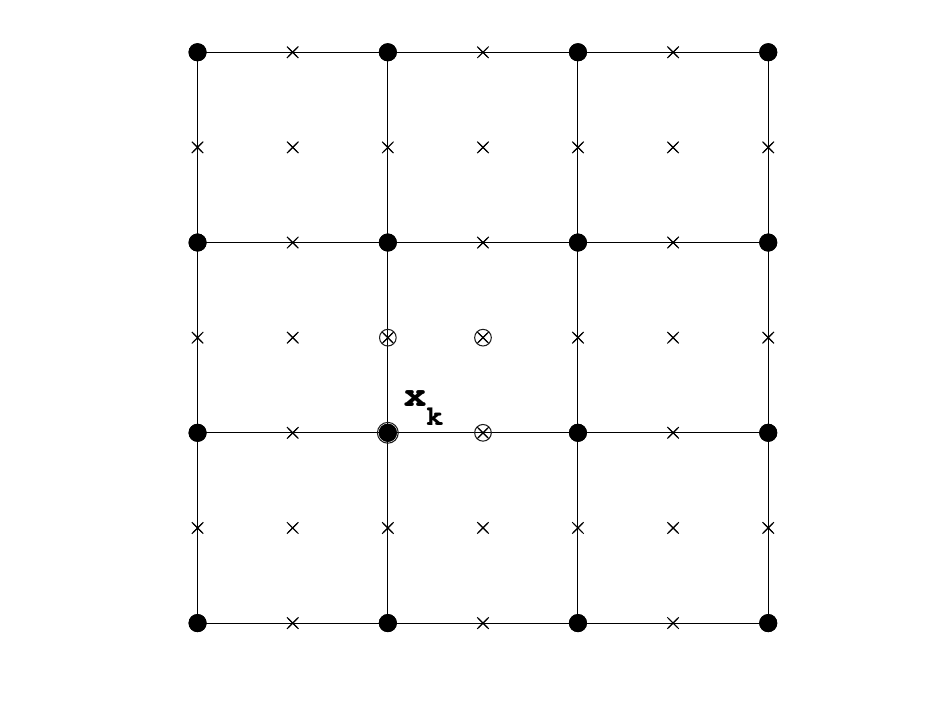}&
    \includegraphics[height=4cm]{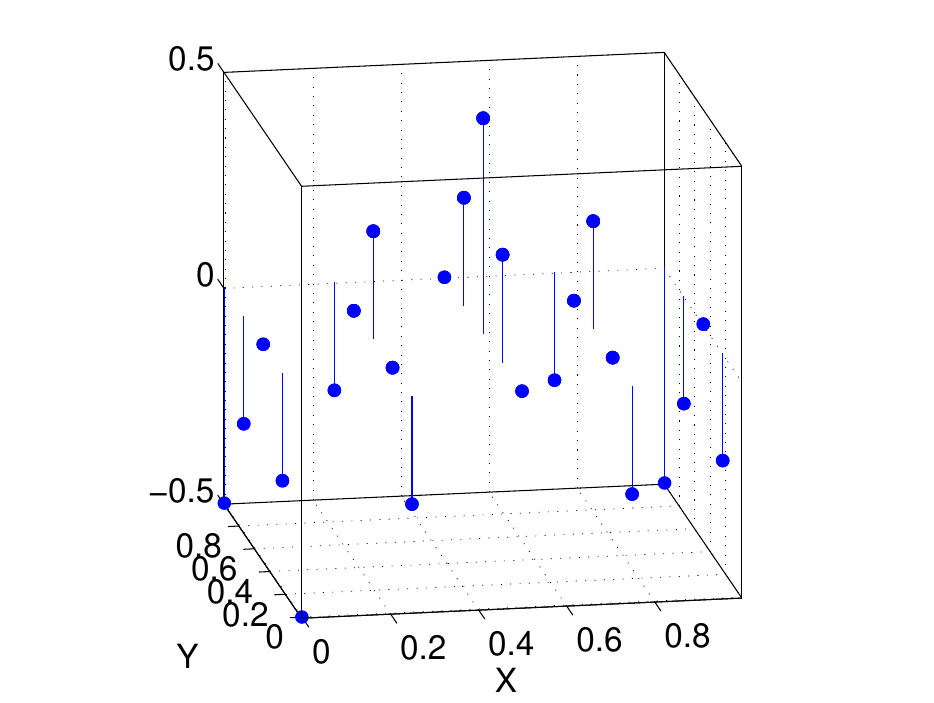}\\
    $(a)$ $2p\times 2p$ neighbors &
    $(b)$ $5\times 5$\\
    \includegraphics[height=4cm]{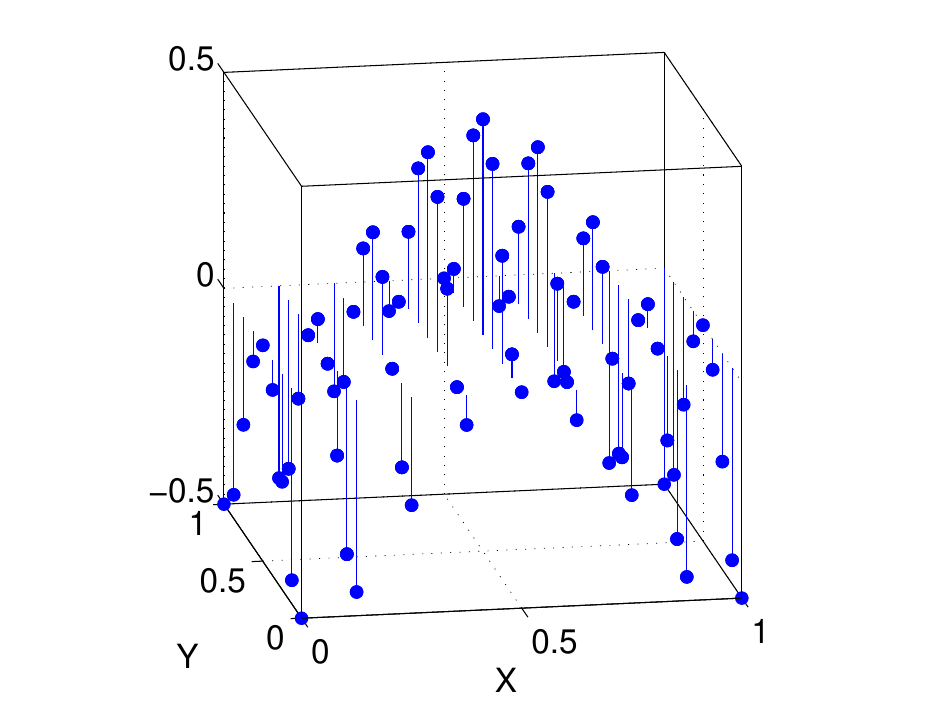}&
    \includegraphics[height=4cm]{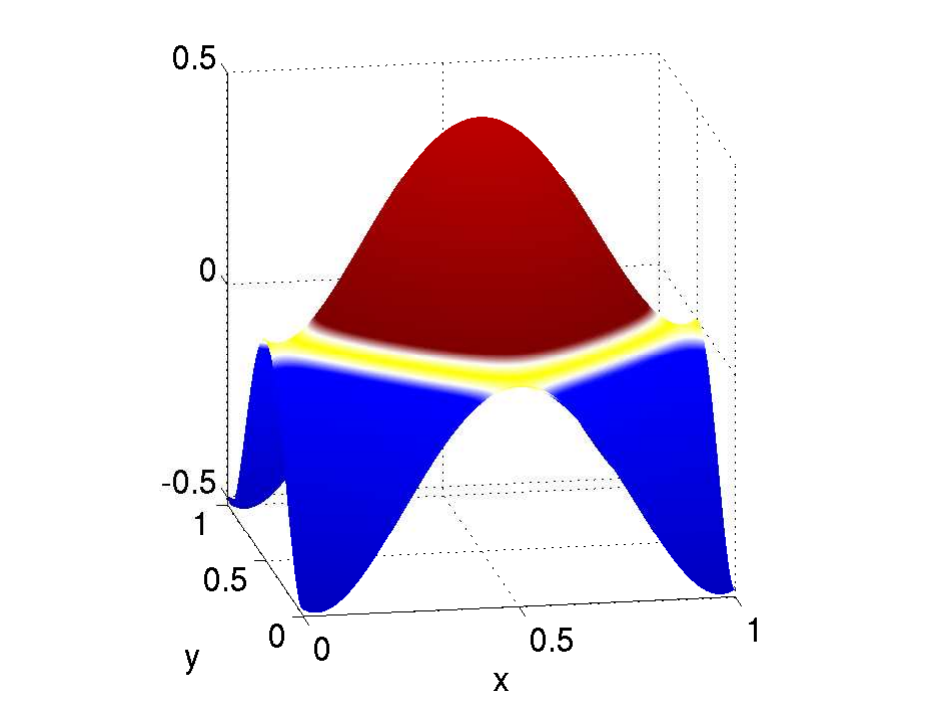}\\
    $(c)$ $9\times 9$ &
    $(d)$ $65\times 65$\\
  \end{tabular}
\end{center}
  \caption{Iterative interpolation and subdivision process. $(a)$~An example of a mesh at $2$ resolutions. The node $\bm x_k\in\mathcal G^s$ and its $2p\times 2p$ neighbors in $\mathcal G^s$ for $p=2$ are marked with $\bullet$. Nodes in $\mathcal G^{s+1}$ those are not present in $\mathcal G^s$ are marked with $\small\times$, and among them, $\bm x_{2k+1}$ are marked with $\otimes$. $\mathcal P_k(\bm x)$ takes a value $1$ on $\bm x_k$ and $0$ on all other nodes $\bullet$, and is used to interpolate new values on three $\otimes$ nodes. We start with $\varphi(\bm x) = c_k$ on all $\bullet$ nodes, and evaluate $\varphi(\bm x) = \sum_k c_k\mathcal P_k(\bm x)$ on all $\times$ nodes, thereby resulting into $\varphi(\bm x)$ on $\mathcal G^{s+1}$. $(b)$ The initial data on $5\times 5$~nodes. $(c)$ Interpolated data on $9\times 9$ nodes. $(d)$ The resulting function with $65\times 65$ nodes.}
  \label{fig:ngh}
\end{figure}

On a given two-dimensional mesh $\mathcal G^s$, the starting point for the iterative interpolation is a function evaluation $\{c_k\}$ at each node $\bm x_k$. This mesh $\mathcal G^s$ is refined dyadically to form a new mesh $\mathcal G^{s+1}$. Fig~\ref{fig:ngh}($a$) uses $\bullet$ and $\times$ to denote nodes in $\mathcal G^s$ and $\mathcal G^{s+1}\backslash\mathcal G^s$, respectively. The given sample $\{c_k\}$ on $\bullet$ nodes is interpolated to $\times$ nodes, and a new sample is obtained in the mesh $\mathcal G^{s+1}$, which has been illustrated in Fig~\ref{fig:ngh}. The process can be repeated until $s\rightarrow\infty$. The iterative interpolation extends $\{c_k\}$ to a function $\varphi(\bm x)$ on the entire domain $\Omega$~\cite{Dubuc89,Dubuc95}. Using a one-dimensional mesh, \citet{Dubuc89} proved that if one assigns $c_k=1$ on a specific node $x_k$ and $c_k=0$ on all other nodes of a given mesh, and the mesh is refined, then the iterative interpolation of this data $\{c_k\}$ results into a uniformly continuous fundamental function. \citet{Dubuc95} studied the two-dimensional iterative interpolation. The resulting fundamental function depends on the specific interpolation (see,~\cite{Donoho96}). 

First, at each node $\bm x_k\in\mathcal G^s$, $\varphi(\bm x)=\sum\limits_{k}c_k\mathcal P_k(\bm x)$ is built by constructing a local polynomial
$$
\mathcal P_k(\bm x_l)=\left\{
    \begin{array}{ll}
      1 &\hbox{ if } \bm x_l = \bm x_k\\
      0 &\hbox{ if } \bm x_l \ne \bm x_k
    \end{array}
    \right.
$$
using $2p\times 2p$ neighbors of the node $\bm x_k$ such that $\varphi(\bm x_k) = c_k$. Fig~\ref{fig:ngh}($a$) presents $2p\times 2p$ neighbors of the node $\bm x_k$ on the coarse mesh, and its neighbors $\bm x_{2k+1}$ on the refined mesh. Next, $\varphi(\bm x)$ is extended to the mesh $\mathcal G^{s+1}$ by 
assigning the value of $\varphi(\bm x_k)$ to $\varphi(\bm x_{2k})$ on $\bullet$ nodes, and using $\varphi(\bm x_{2k+1})=\mathcal P_k(\bm x_{2k+1})$ on $\times$ nodes. As a result, we have $\varphi(\bm x_k)$ for all nodes $\bm x_k\in\mathcal G^{s+1}$. The interpolation is iterated on $\mathcal G^{s+1}$ to extend $\varphi(\bm x)$ on the mesh $\mathcal G^{s+2}$.  Clearly, one obtains $\varphi(\bm x)$ for every $\bm x\in\Omega$ by employing the interpolation and the subdivision repeatedly. The limit function $\varphi(\bm x)$ of this iterative interpolation is a two-dimensional fundamental function~\cite{Dubuc95}. 

\begin{figure}
  \centering
  \begin{tabular}{cc}
    \includegraphics[height=3.5cm]{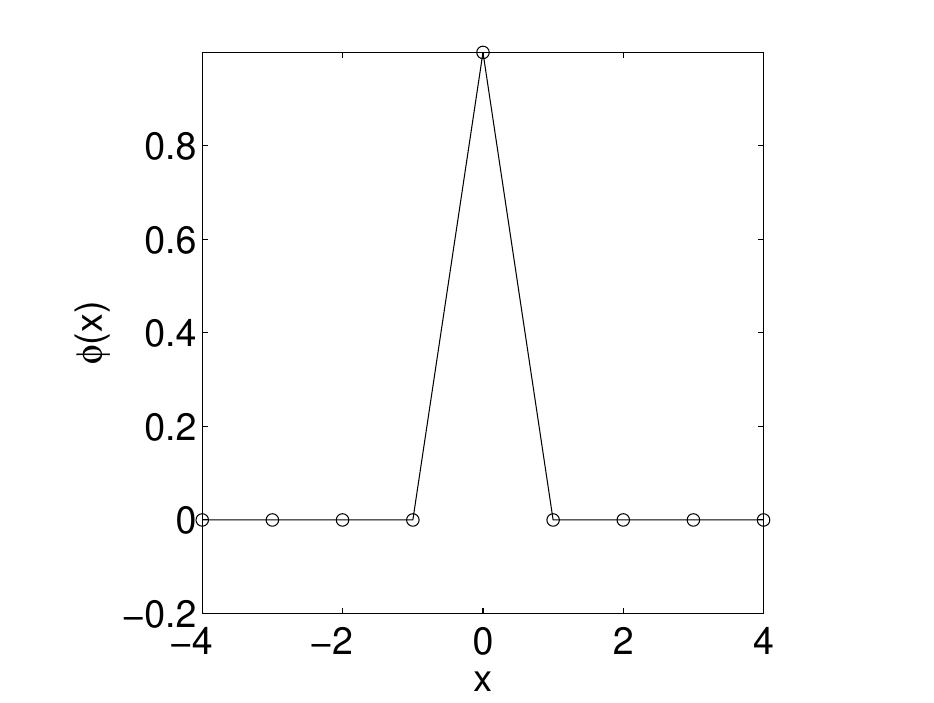}
    &
    \includegraphics[height=3.5cm]{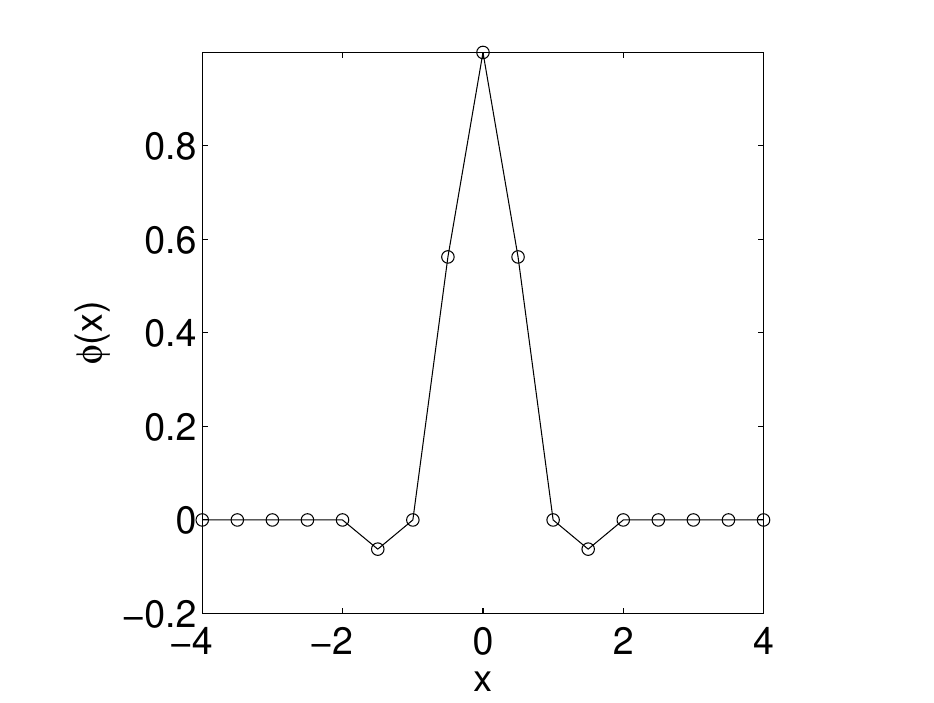}
    \\
    iteration, $s=0$ & iteration, $s=1$ \\
    \includegraphics[height=3.5cm]{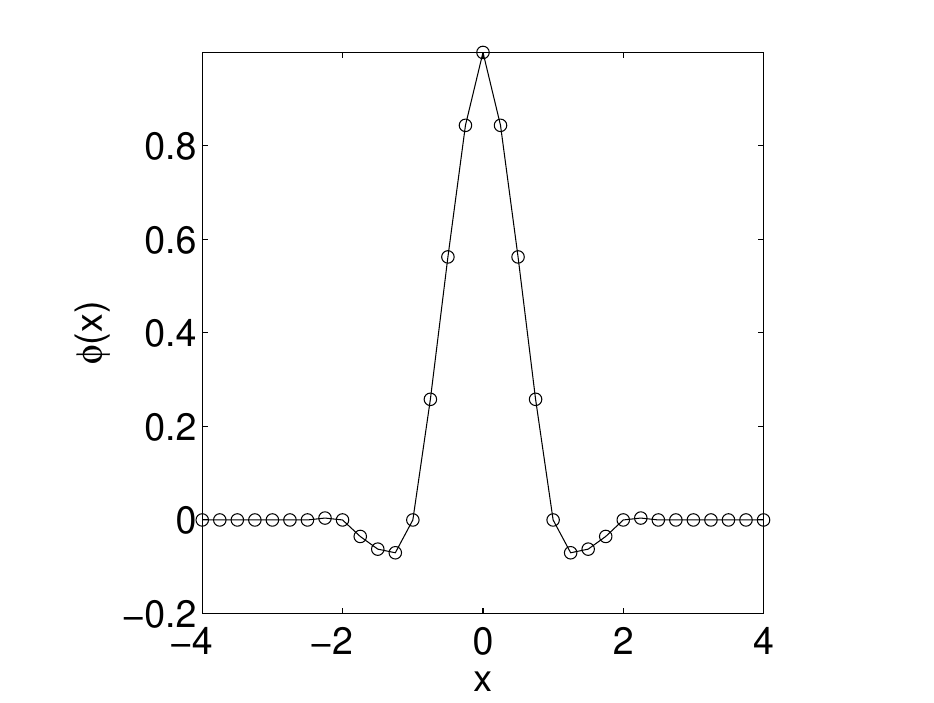}
    &
    \includegraphics[height=3.5cm]{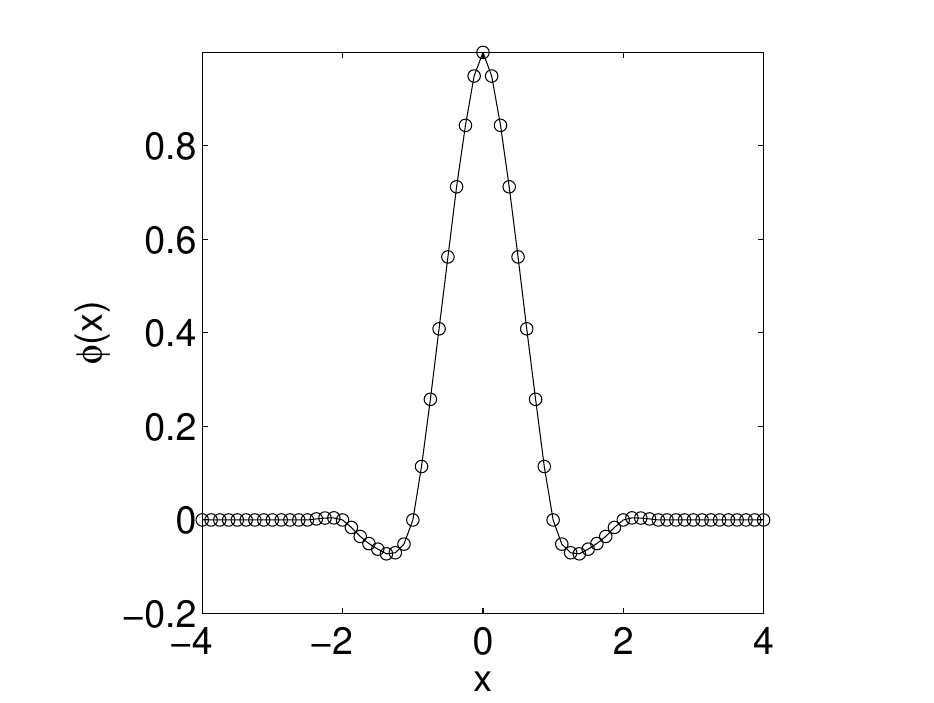}
    \\
    iteration, $s=2$ & iteration, $s=3$ \\
    \includegraphics[height=3.5cm]{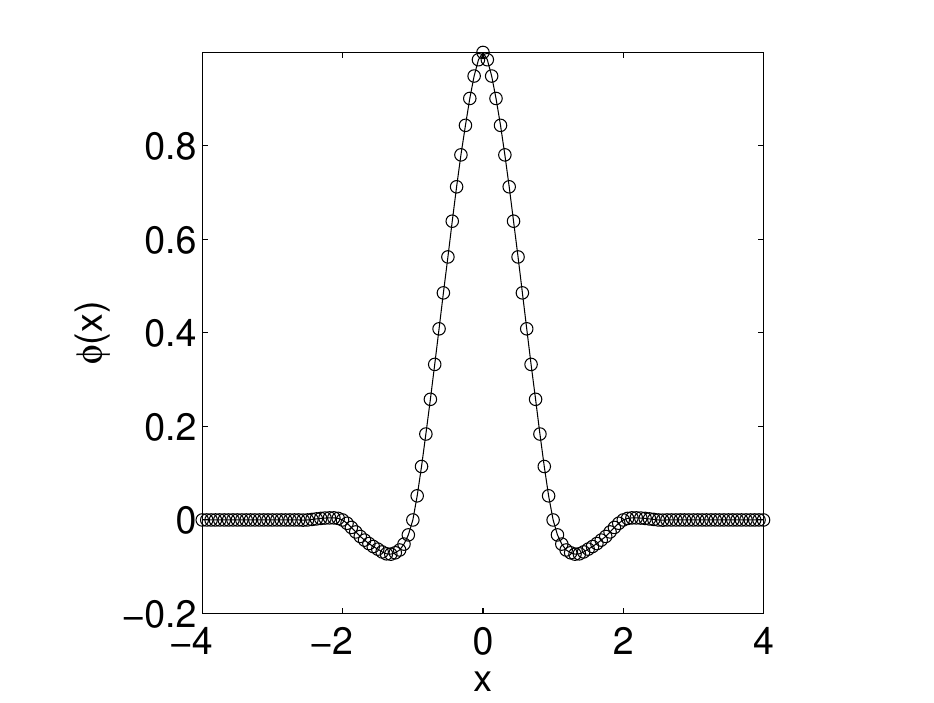}
    &
    \includegraphics[height=3.5cm]{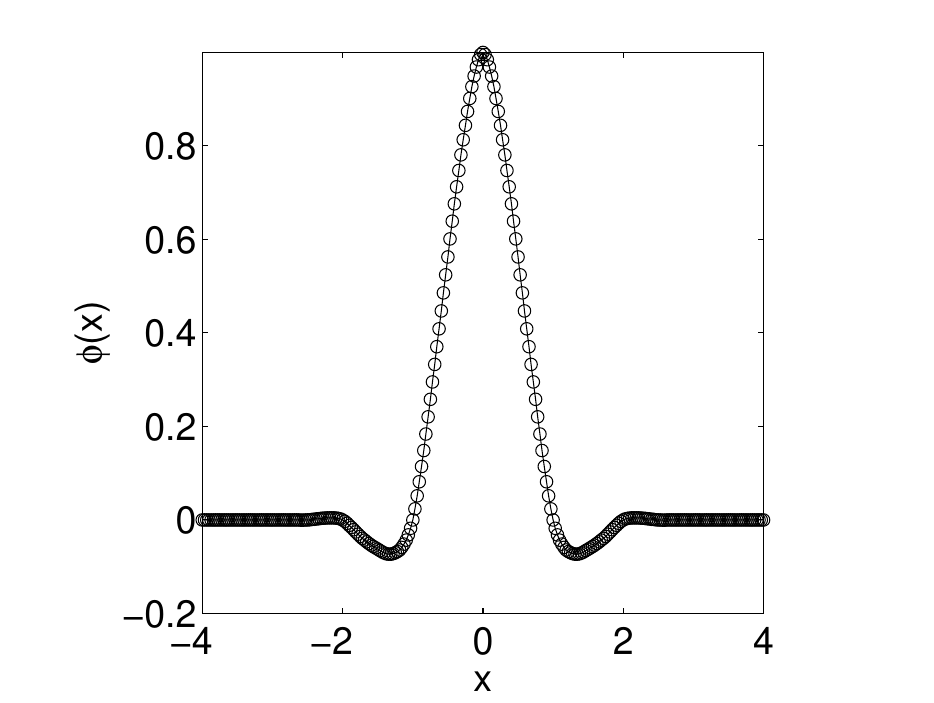}
    \\
    iteration, $s=4$ & iteration, $s=5$ 
  \end{tabular}
  \caption{A one-dimensional fundamental function generated with a cubic interpolation process. We see that the interpolation tends to a unique function.}
  \label{fig:phi1d}
\end{figure}

We now present two examples. In the first example, consider the nodes $x=\{-4,-3,-2,-1,0,1,2,3,4\}$ and the data $c=\{0,0,0,0,1,0,0,0,0\}$, where $c=\varphi(x)$ has been plotted in Fig~\ref{fig:phi1d}($a$). Using interpolation with a cubic polynomial, {\em i.e.} with $4$ data points, we illustrate the sequence of refinements and corresponding $\varphi(x)$ at each iteration in Figs~\ref{fig:phi1d}$(b-f)$. This shows how to construct the fundamental function. 
The second example presents the approaching function of a 2D interpolation. Fig~\ref{fig:ngh}($b$) shows a sampled data $\{c_k\}$ on a $5\times 5$ initial mesh. This data has been interpolated with $p=2$, {\em i.e.} with a $4\times 4$ stencil, to form a new $9\times 9$ sample $\{c_k\}$ ({\em e.g.} Fig~\ref{fig:ngh}($c$)). As marked in Fig~\ref{fig:ngh}$(a)$, based on $4\times 4$ $\bullet$ neighbors of the node $\bm x_k$, interpolation is done on three associated $\tiny\otimes$ nodes, which is repeated for each node of the initial $5\times 5$ data. We now refine the $9\times 9$ mesh, and repeat the interpolation on each refined mesh with $p=2$, which we have stopped on a $65\times 65$ mesh, for this example. Fig~\ref{fig:ngh}($d$) shows the constructed function on the $65\times 65$ mesh.

These numerical illustrations demonstrate that iterative interpolation of a given data set converges to a continuous function $\varphi(\bm x)$. In the next section, we present some beautiful properties of $\varphi(\bm x)$. 

%
%

\subsection{A basis for discretization}
This section presents the construction of a basis for the space $\mathcal V^s$ so that we can define the trial solution~(\ref{eq:trl}). The space $\mathcal V^s$ is a collection of linear combinations of its basis. We call each member of the basis a scaling function associated with a corresponding mesh $\mathcal G^s$ that is a collection of rectangles~\cite{Mallat2009,Sweldens97}. In 1D, the fundamental function in Fig~\ref{fig:phi1d} is translated to form a linearly independent set of scaling functions. 
For example, at each node $\bm x_k$ of $\mathcal G^s$, we associate a scaling function $\varphi_k(\bm x) = \varphi(\bm x-\bm x_k)$ based on the dyadic interpolation of order $p$, and as a result,  the basis $\left\{\varphi_k(\bm x)\right\}$ of $\mathcal V^s$ is formed. 

\begin{figure}
  \centering
  \begin{tabular}{cc}
    \includegraphics[height=3.5cm]{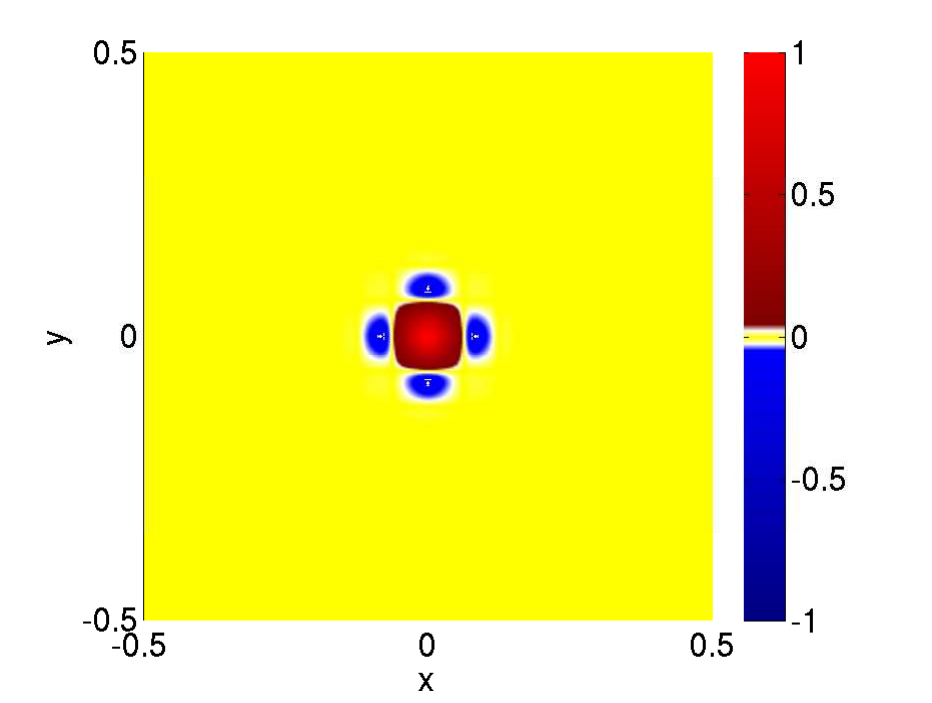}&
    \includegraphics[height=3.5cm]{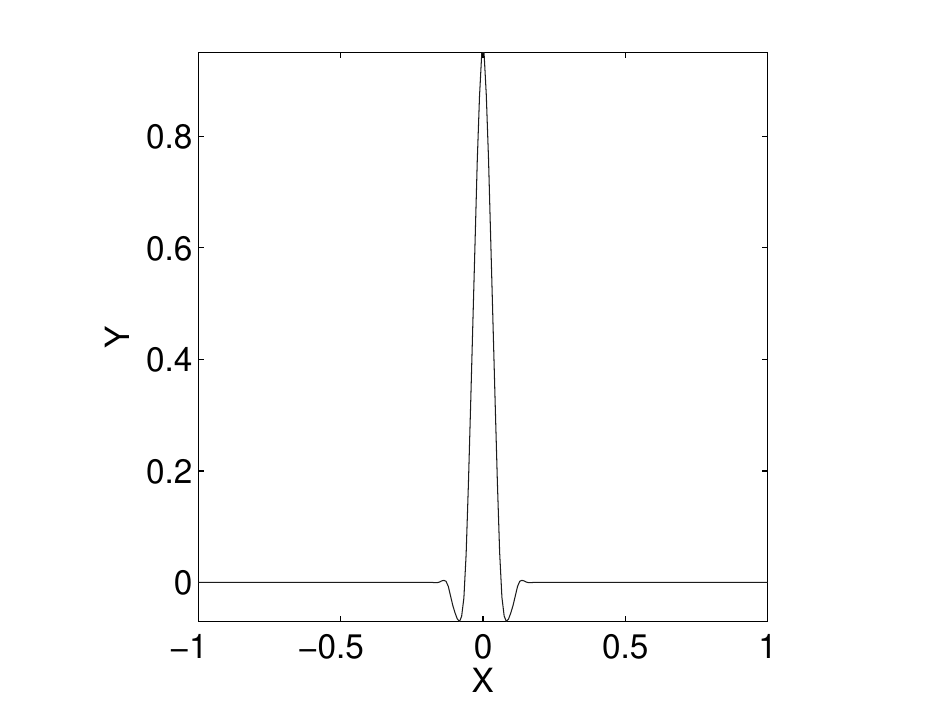}\\
    \includegraphics[height=3.5cm]{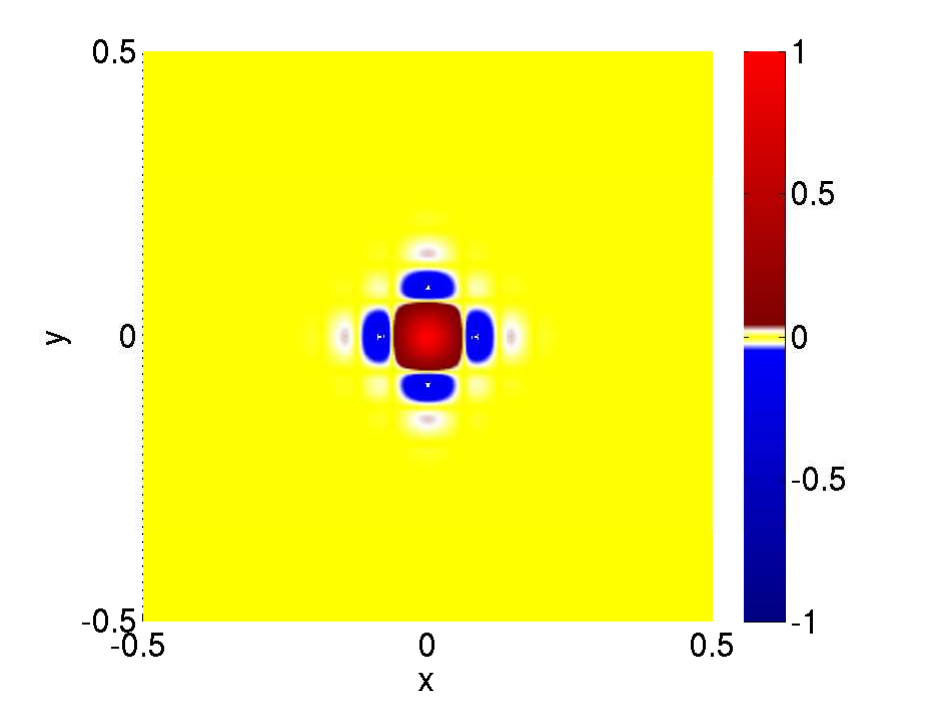}&
    \includegraphics[height=3.5cm]{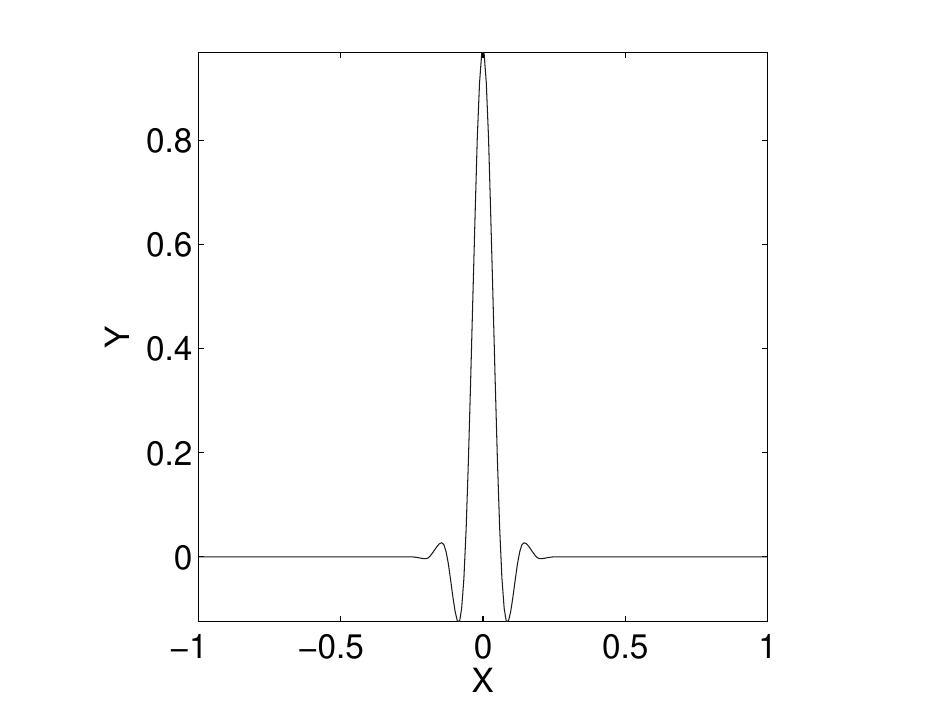}\\
    \includegraphics[height=3.5cm]{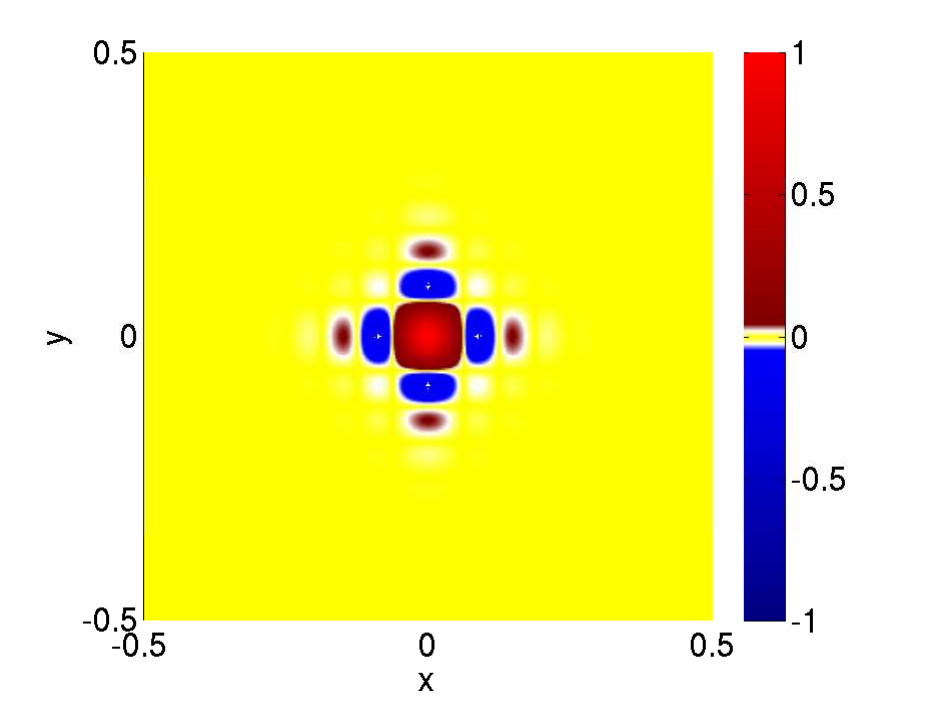}&
    \includegraphics[height=3.5cm]{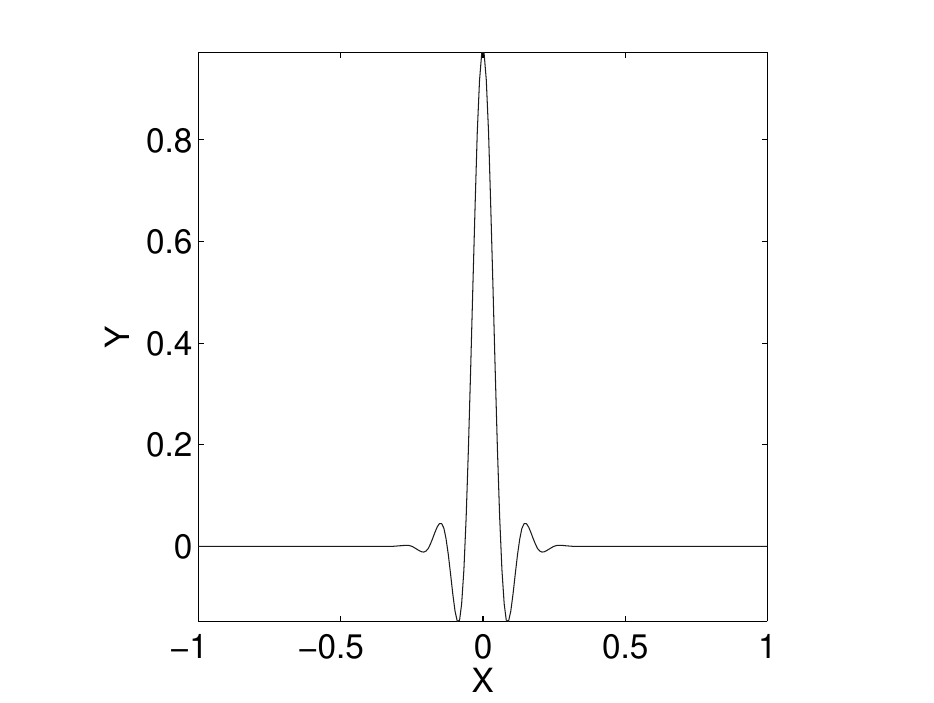}\\
    \includegraphics[height=3.5cm]{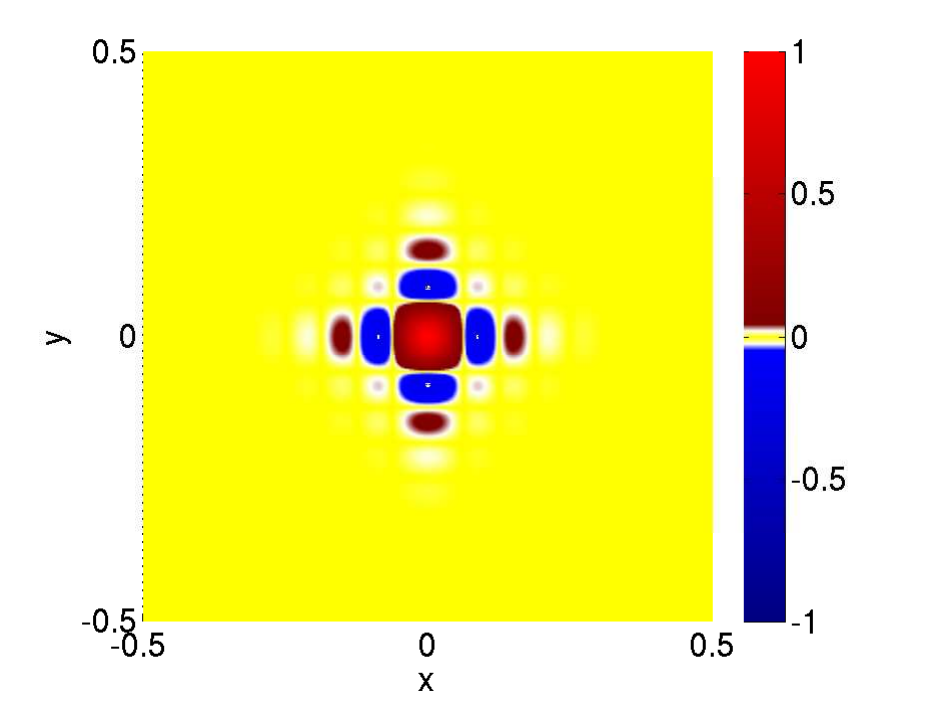}&
    \includegraphics[height=3.5cm]{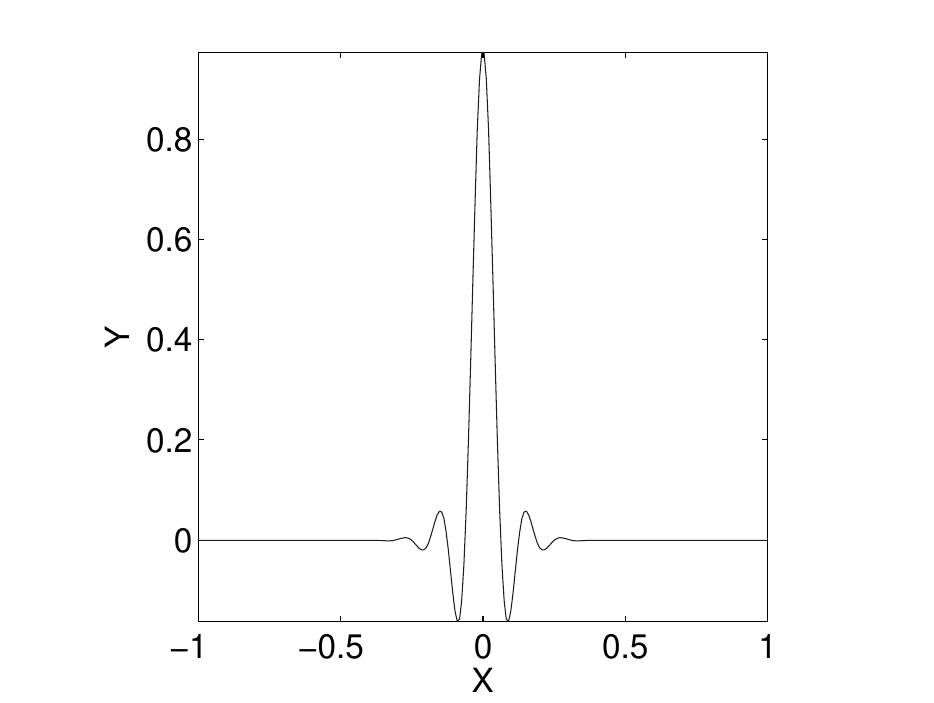}
  \end{tabular}
  \caption{Scaling function $\varphi(x,y)$ in the domain $[-1,1]\times[-1,1]$. In the left column, only the portion $[-0.5,0.5]\times[-0.5,0.5]$ of the domain is shown for clarity. Vertically downward, rows correspond to $p=2$, $4$, $6$, and $8$, respectively. In the right column, $\varphi(x,0)$ is shown for each $p$. Each curve in the right column has exactly $2p$ zeros; however, at large $p$, fluctuation of $\varphi(x,0)$ away from the center is not visible with naked eye.}
  \label{fig:phi}
\end{figure}

We now demonstrate a few examples of constructed scaling functions $\varphi(\bm x)$ in the square $[-1,1]\times[-1,1]$. They are presented in Fig~\ref{fig:phi} for $p=2,\,4,\,6,\,\hbox{ and } 8$. For each $p$, $\varphi(x,0)$ is also shown. Each of these two-dimensional scaling functions, $\varphi(\bm x)$, is symmetric with respect to $x=0$, $y=0$, and $y=\pm x$. Note that the exact mathematical form of $\varphi(\bm x)$ may not be known. We only need to know its initial function evaluation $\{c_k\}$, and the interpolation process. 

In order to have an equivalent resolution, we assigned $\varphi(x,y)=1$ on $(x,y)=(0,0)$, and $\varphi(x,y)=0$ on all other nodes in a $33\times 33$ mesh of the square $[-1,1]\times[-1,1]$. As can be seen from Fig~\ref{fig:phi}, the support of $\varphi(x,y)$ increases with $p$. For $p=8$, $\varphi(x,y)$ vanishes for all $(x,y)\notin(-15\Delta x,15\Delta x)\times(-15\Delta y, 15\Delta y)$, where $\Delta x=2/32$ and $\Delta y=2/32$.

A restriction of the two-dimensional scaling function $\varphi(\bm x)$ along a line that is parallel to a coordinate axis is exactly the fundamental function $\varphi(x)$ of~\citet{Dubuc89}, which has the following properties.
\begin{itemize}
\item $\varphi(x)$ is an interpolating polynomial, which vanishes outside the interval $[x_{-2p+1}, x_{2p-1}]$, where $p$ is an integer. Moreover, $\varphi(0) = 1$ and $\varphi(x)$ has exactly $4p-2$ zeros in the interval $[x_{-2p+1},x_{2p-1}]$.

\item $\varphi(x)$ is symmetric about $x=0$; {\em i.e.}, $\varphi(x)$ is an even polynomial.

\item $\varphi(x)$ is uniformly continuous for all $p$ on any finite interval, and is differentiable for $p>1$. Moreover, $\varphi(x)$ has at least two continuous derivatives for $p=3$ (see,~\cite{Dubuc89}). 

\item A linearly independent set $\{\varphi_k(x)\}$ is obtained from translations of $\varphi(x-x_k)$, which satisfies $\varphi_k(x_j)=\delta_{kj}$. Such a basis $\{\varphi_k(x)\}$ reproduces polynomials up to degree $2p-1$, which is an important property for developing efficient numerical methods. 

%
\end{itemize}

According to~(\ref{eq:trl}), the discretization of $P(\bm x)$ is a collection of continuous functions $\varphi_k(\bm x)$, which can be differentiated to approximate the derivatives of $P(\bm x)$. 

\subsection{Differentiation}
This section studies the weighted residual collocation method for the numerical differentiation of the trial solution~(\ref{eq:trl}) such that $\langle\frac{\partial}{\partial x}P^{\mathcal N}(\bm x),\delta(\bm x-\bm x_k)\rangle=0$ on a mesh $\mathcal G^s$. The present collocation method uses some basic properties of $\varphi_k(\bm x)$, and considers a corresponding expansion of $\frac{\partial}{\partial x}P^{\mathcal N}(\bm x)$ in the form of~(\ref{eq:trl}); {\em i.e.}
\begin{equation}
  \label{eq:trld}
  \frac{\partial}{\partial x}P^{\mathcal N}(\bm x) \equiv \sum_{k=0}^{\mathcal N-1}c'_k\varphi_k(\bm x) = \sum_{k=0}^{\mathcal N-1}c_k\frac{\partial}{\partial x}\varphi_k(\bm x),
\end{equation}
where $c'_k$'s denote expansion coefficients for the first derivative. According to~\citet{Dubuc89}, the middle part of~(\ref{eq:trld}) states that the derivative $\frac{\partial}{\partial x}P^{\mathcal N}(\bm x)$ is a uniformly continuous function in the domain $\Omega$, which is obtained by the DD subdivision. The last part of~(\ref{eq:trld}) states that the expansion coefficients, $c'_k$, are obtained by the exact derivative of $\varphi_k(\bm x)$, where $c_k$ are already known. Let us simplify~(\ref{eq:trld}) using some fundamental properties of $\varphi_k(\bm x)$.  

Since $\varphi_k(x)$ (for fixed $y$) is an even function with respect to $x=x_k$, and $\varphi_k(x)$ has exactly $4p-2$ zeros within its support $[x_{k-2p+1},x_{k+2p-1}]$, the following statements are true. $(i)$~The $1$st derivative $\varphi_k'(x)$ is an odd function, $(ii)$~it vanishes at $x_k$, {\em i.e.} $\varphi'_k(x_k)=0$, $(iii)$~$\varphi'_k(x)$ takes nonzero values at zeros of $\varphi_k(x)$ in $(x_{k-2p+1},x_{k+2p-1})$, and $\varphi_k'(x)$ vanishes for all other $x\notin(x_{k-2p+1},x_{k+2p-1})$ (see, \cite{Dubuc89}).

Using these properties, it is easy to see that combining $\langle\frac{\partial}{\partial x}P^{\mathcal N-1}(\bm x),\,\tilde\varphi(\bm x)\rangle=0$ with eq~(\ref{eq:trld}) results in
\begin{equation}
  \label{eq:trld1}
  \frac{\partial}{\partial x}P^{\mathcal N}(x_j)\equiv \sum_{k=0}^{\mathcal N-1}c'_k\varphi_k( x_j)  = \sum\limits_{k=j-2p+1}^{j+2p-1}c_k\varphi_k'(x_j),
\end{equation}
where it is convenient to use a one-dimensional notation since the derivative is taken for a fixed $y$. 
To evaluate $\varphi'_k(x_j)$ on the right side of~(\ref{eq:trld1}), let us obtain $\varphi_k(x)$ 
from the interpolation process, without knowing the actual mathematical form of $\varphi_k(x)$, using the barycentric formula~(see~\cite{Dubuc89,Winrich69,Berrut2004}) 
$$
\varphi_k(x) = \frac{w_k(x)}{\sum\limits_{l=k-2p+1}^{k+2p-1}w_l(x)}.
$$
at $x_j$ for $j=k-2p+1,\ldots,k+2p-1$. The weights $w_k(x)$ are associated with $2p+1$ nodes, and are extended from the iterative interpolation process that derives $\varphi(x)$. 
In order to employ the weighted residual collocation method, let us define,
\begin{equation}
  \label{eq:wgh}
  \frac{1}{w_k(x)} = (x-x_k)\prod\limits_{j\ne k}(x_k-x_j)
\end{equation}
and
$$
s(x) = \sum\limits_lw_l(x)(x-x_j)
$$
and assume the weighted inner product
$$
\left\langle[\varphi_k(x)s(x)]',\tilde\varphi_k(x)\right\rangle = 0.
$$
A quick calculation leads to
$$
\varphi_k'(x_j) = \left\{
  \begin{array}{cc}
    \frac{w_k(x_j)}{w_j(x_k)(x_k-x_j)}&\hbox{ for } k\ne j\\
    -\sum\limits_{k\ne j}\varphi_k'(x_j)&\hbox{ for } k= j.
  \end{array}
\right.
$$
Clearly, knowing the ingredients, $w_k$'s, of the iterative interpolation, we are able to compute derivatives of $\varphi_k(x)$ exactly on all nodes. Using a equally spaced one-dimensional dyadic mesh with $p=2,\,3$, we find that the values $\varphi'_k(x_j)$ obtained from the above formula agrees exactly with those presented by~\citet{Dubuc89}. 

Using the above expression for $\varphi'_k(x_j)$, (\ref{eq:trld1}) provides the first derivative of the trial solution~(\ref{eq:trl}) at all nodes. It is also clear from~(\ref{eq:trld1}) that the first derivative of the trial solution is given by the products of its nodal values with $\varphi'_k(x_j)$'s. Clearly, the process has $\mathcal O(\mathcal N)$ complexity, which does not require global operations on the $\mathcal N\times\mathcal N$ differentiation matrix. In other words, one does not need to store the differentiation matrix explicitly, and  the overall CPU time for the discretization is asymptotically optimal if $\mathcal N\rightarrow\infty$.

The weighted residual collocation method for computing the second order derivative of the trial solution~(\ref{eq:trl}) is computed using the nodal values $\varphi'_k(x_j)$.
Let us denote $c'_k = \frac{\partial}{\partial x}P^{\mathcal N}(x_k)$, and rewrite~(\ref{eq:trld1}) for the second derivative,
\begin{equation}
  \label{eq:trld2}
  \frac{\partial^2}{\partial x^2}P^{\mathcal N}(x_j) = \sum\limits_{k=j-2p+1}^{j+2p-1}c'_k\varphi_k'(x_j).
\end{equation}
Since $\varphi'_k(x_j)$'s have been computed, $c'_k$'s can be computed from~(\ref{eq:trld1}), and hence, the right side of~(\ref{eq:trld2}) can be evaluated. However, we can also rewrite the right side of~(\ref{eq:trld2}) as
$$
  \sum\limits_{k=j-2p+1}^{j+2p-1}c'_k\varphi_k'(x_j)
  =
  \sum\limits_{k=j-2p+1}^{j+2p-1}c_k\varphi''_k(x_j),
$$
where $\varphi''_k(x_j)$'s are some necessary weights for the second derivative.
Using a similar approach, the weights $\varphi''_k(x_j)$ for the second order derivative of the trial solution~(\ref{eq:trl}) are given in terms of the first derivative of the scaling function by
$$
\varphi_k''(x_j) = \left\{
  \begin{array}{cc}
    -2\varphi_k'(x_j)\left[\sum\limits_{i\ne k}\varphi_k'(x_i)-\frac{1}{x_j-x_k}\right] & \hbox{ for } k\ne j\\
    -\sum\limits_{k\ne j}\varphi_k''(x_j)&\hbox{ for } k= j.
  \end{array}
\right.
$$
For a dyadic interpolation with $p=3$, $\varphi_k(x)$ is twice differentiable. We have checked that eq.~(\ref{eq:trld2}) provides nodal values of the second derivatives of $\varphi_k(x)$, which agree exactly with those derived by~\citet{Dubuc89}.

\Add{On a two-dimensional mesh, global operations and direct solvers lead to $\mathcal O(\mathcal N^3)$ complexity, and hence, extremely high memory and CPU time~(see,~\cite{Kannan2009}). The present methodology employs local operations on a node  ($\bm x_k$) in each direction to discretize a PDE. This is an important computational benefit, which comes from the weighted residual collocation method~({\em e.g.}~\cite{Bruce}).  According to~(\ref{eq:trld}), the leading order error of the differentiation on a mesh can be shown $\mathcal O(\Delta x^{2p})$ as $\Delta x\rightarrow 0$ for a fixed $p$. Clearly, for a fixed $\Delta x$ and $\varepsilon=\mathcal O(\Delta x^2)$ at $p=1$, the error decreases like $\mathcal O(\varepsilon^p)$  as $p\rightarrow\infty$. The following example verifies this estimate.}
\subsubsection{Example\\}\label{ex}

\begin{figure}
  \centering
  \includegraphics[height=6cm]{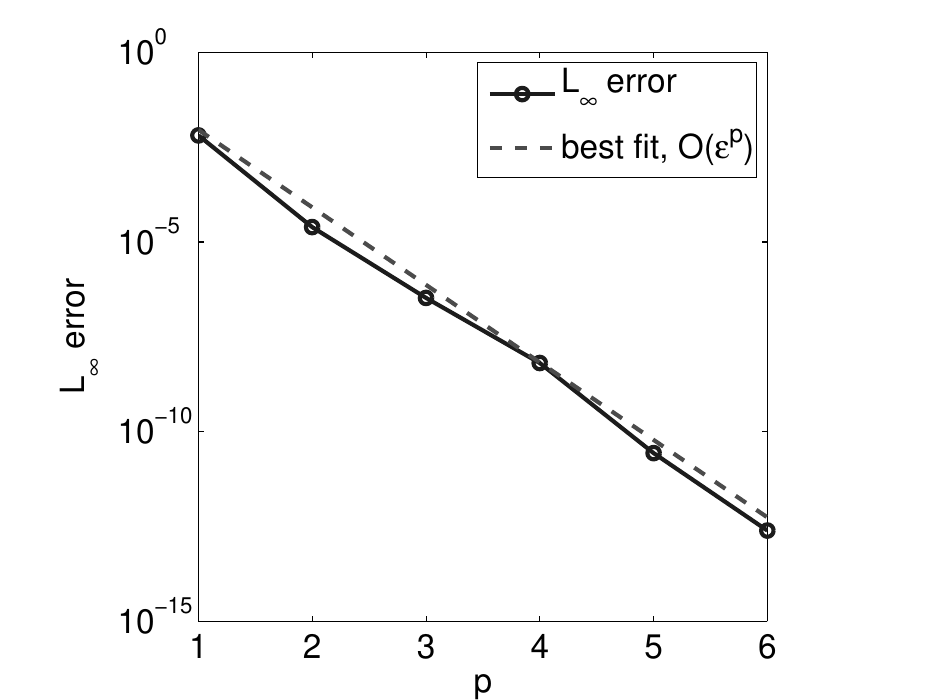}
  \caption{For the example~(\ref{ex}), a verification of $\mathcal O(\varepsilon^p)$ for the $L_{\infty}$ error as a function of $p$, where $\varepsilon$ represents the error at $p=1$. }
  \label{fig:laperr}
\end{figure}

Consider the function
$$
u(x,y) = \frac{1}{\pi\nu}\exp\left(-\frac{x^2+y^2}{\nu}\right),\quad (x,y)\in[-1,1]\times[-1,1]
$$
for which $\nabla^2u$ is known exactly, and we have used this function to check the numerical error. Note, depending on the value of $\nu$, this function has a singularity and a localized structure near the origin $(0,0)$.   For $\nu =10^{-2}$, we have estimated the error $|\nabla^2u(x,y)-\nabla^2u^{\mathcal N}(x,y)|_{\infty}$ on a $129\times 129$ mesh with $p=1$, $2$, $3$, $4$, $5$, and $6$, and the results are listed in the Table~\ref{tab:ex1ep}. As expected, the error is reduced with increased order~($p$) of interpolation. \Add{Fig~\ref{fig:laperr} shows that the data in Table~\ref{tab:ex1ep} follow $\varepsilon^p$.}  This behavior of the error is also consistent with the error bounds given by~\citet{Dubuc89}. 
\begin{table}
  \centering
  \begin{tabular}{|c|c|c|}
    \hline
     $p$& degree of $\varphi(\bm x),\,2p-1$& $|\nabla^2u(x,y)-\nabla^2u^{\mathcal N}(x,y)|_{\infty}$\\
    \hline
    $1$& $1 $& $6.37\times 10^{-3}$ \\
    $2$& $3 $& $2.46\times 10^{-5}$ \\
    $3$& $5 $& $3.35\times 10^{-7}$ \\
    $4$& $7 $& $6.46\times 10^{-9}$ \\
    $5$& $9 $& $2.70\times 10^{-11}$ \\
    $6$& $11$& $2.46\times 10^{-13}$\\
    \hline
  \end{tabular}
  \caption{The errors $|\nabla^2u(x,y)-\nabla^2u^{\mathcal N}(x,y)|_{\infty}$ for estimating the Laplacian of the function in example~(\ref{ex}) with $p=1$, $2$, $3$, $4$, $5$, and $6$ have been listed.}
  \label{tab:ex1ep}
\end{table}

\section{Numerical simulation and verification}\label{sec:pde}
In this section, we examine the proposed spatial discretization methodology with a few representative examples, where a Krylov method has been used to solve the discrete system~(see,~\cite{Wesseling2000}). More specifically, we have employed the restarted GMRES~(generalized minimal residual) algorithm~\cite{Brown94,Wesseling2000}. First, we examine numerical accuracy with examples where the solution can be derived analytically. Second, we verify the methodology by three representative simulations which are often used as benchmark CFD examples. \Add{For each simulation, $m_x=2$, $m_y=2$, and $p=3$ are used unless it is mentioned otherwise. For example, a mesh with $\mathcal N = 129\times 129$ is obtained with $7$ refinement levels.} 

\subsection{The potential field induced by a swarm of charged colloids}
Electroosmotic flow (EOF) past a swarm of colloidal particles is generated by an externally applied electrical field onto an electrolyte solution ({\em e.g.},~\cite{Alam2002,Alam2012b,Patankar98}). In addition to momentum and Nernst-Planck equations, a model of EOF solves eq~(\ref{eq:bvp}) for the electric potential, $P(x,y)$, that is induced by the local charge distribution, $\rho(x,y)$.

\begin{figure}
  \centering
  \begin{tabular}{cc}
    \includegraphics[height=4cm]{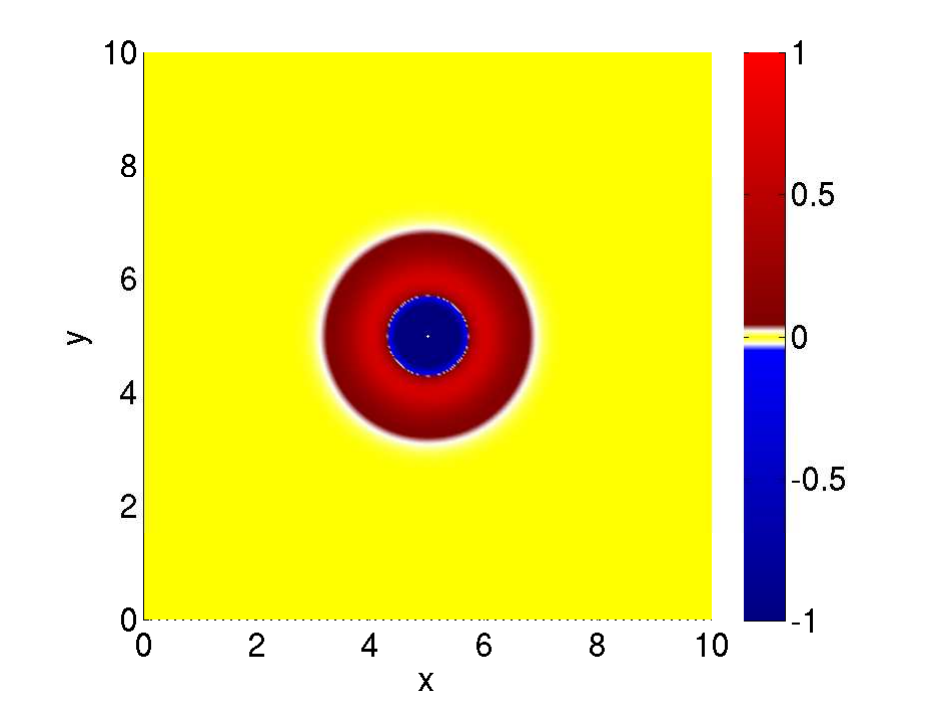}&
    \includegraphics[height=4cm]{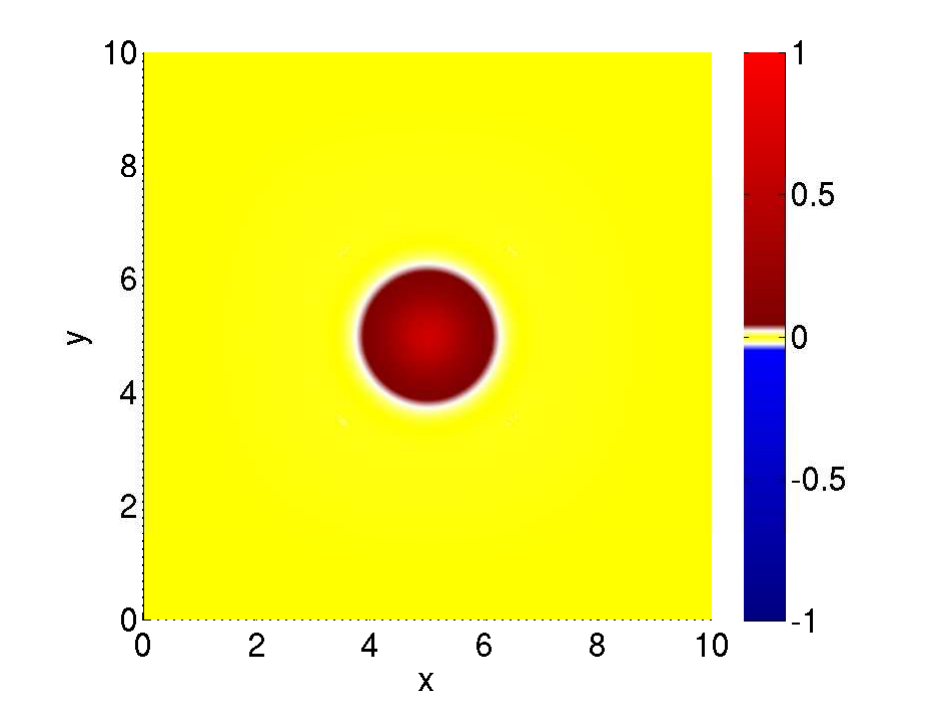}\\
    $(a)$ & $(b)$\\
     \multicolumn{2}{c}{
       \includegraphics[height=4cm]{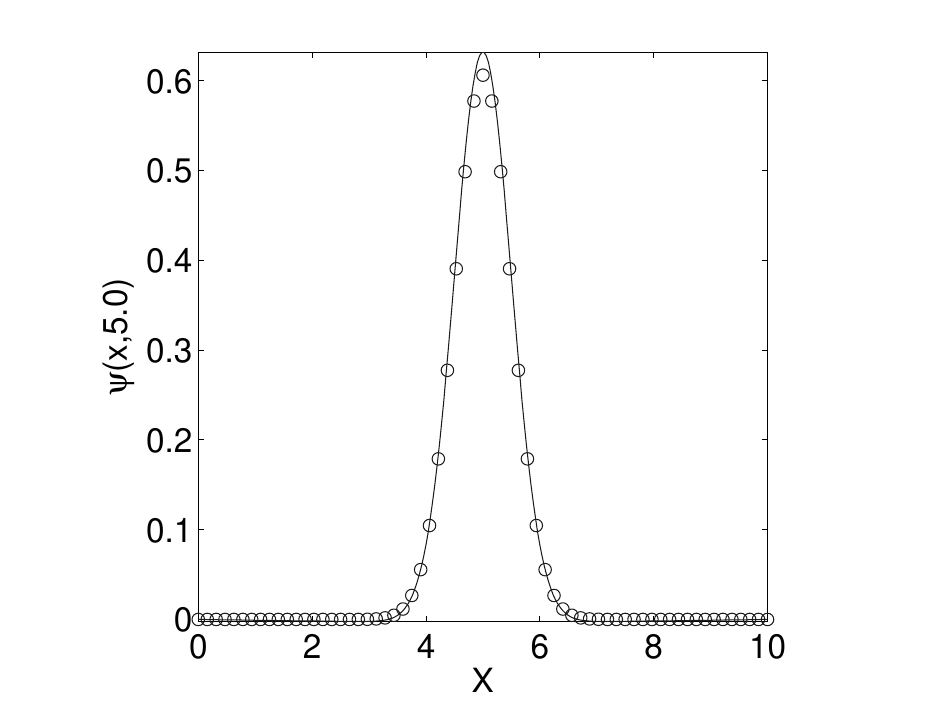}
     }\\
     \multicolumn{2}{c}{
       $(c)$
     }\\
  \end{tabular}
  \caption{`Blue' and `red' color represent a homogeneous swarm of negatively and positively charged particles, respectively. `Yellow' color represents a neutral charge. $(a)$~The prescribed charged distribution is $\rho(x,y)$. $(b)$ The numerical solution of (\ref{eq:bvp}) for $P(x,y)$. $(c)$ $P(x,5.0)$ has been compared with its exact value.}
  \label{fig:vor}
\end{figure}

To verify the present method on simulating the charge induced potential with an idealized EOF, where a manufactured solution is used for a numerical verification purpose, consider the distribution of a negatively charged colloid particles surrounded by a shell of positively charged particles as shown in Fig~\ref{fig:vor}($a$). In the present simulation, the charge distribution has been modelled by
$$
\rho(x,y) = -\frac{4}{\nu_1}\exp(-(x^2+y^2)/\nu_1)\left[1-\frac{x^2+y^2}{\nu_1}\right]
$$
in a domain, $[0,10]\times[0,10]$. This idealized example is useful because we can compare the numerical solution with the exact solution $P_{\hbox{exact}}(x,y) = \exp(-(x^2+y^2)/\nu_1)$ \Add{(e.g. $\nu_1=0.5$)}.
Applying the GMRES method to the discretization of~(\ref{eq:bvp}), and using the above expression for $\rho(x,y)$, we compute the potential $P(x,y)$ on $129\times 129$ uniformly distributed nodes, and the computed potential field is shown in Fig~\ref{fig:vor}$(b)$ as a color filled contour plot. The resolution independent convergence is accepted with respect to the relative residual error $|\nabla^2P(x,y)-\rho(x,y)|_{\infty}/|\rho(x,y)|_{\infty} < 10^{-3}$, and the overall solution does not vary significantly if the mesh is refined. We found that a tolerance~$\le 10^{-4}$ on the relative residual did not have much effect in this case. When the numerical solution $P(x,y)$ was compared with the exact solution in a color filled contour plot, no difference can be identified. Thus, for a more quantitative comparison, we have compared, in Fig~\ref{fig:vor}$(c)$, the numerical potential $P(x,0.5)$ with its corresponding exact solution. An excellent agreement has been observed, where only a negligible error appears near the center of the domain because the potential field has a sharp gradient in that region. 

Although the simulated potential field is too idealized, the present experiment exhibits some usefulness of the methodology to the field of EOF, where high performance CFD techniques are desired.
\subsection{Helmholtz-Hodge decomposition of a vector field}
\begin{figure}
  \centering
  \begin{tabular}{cc}
    \includegraphics[height=4cm]{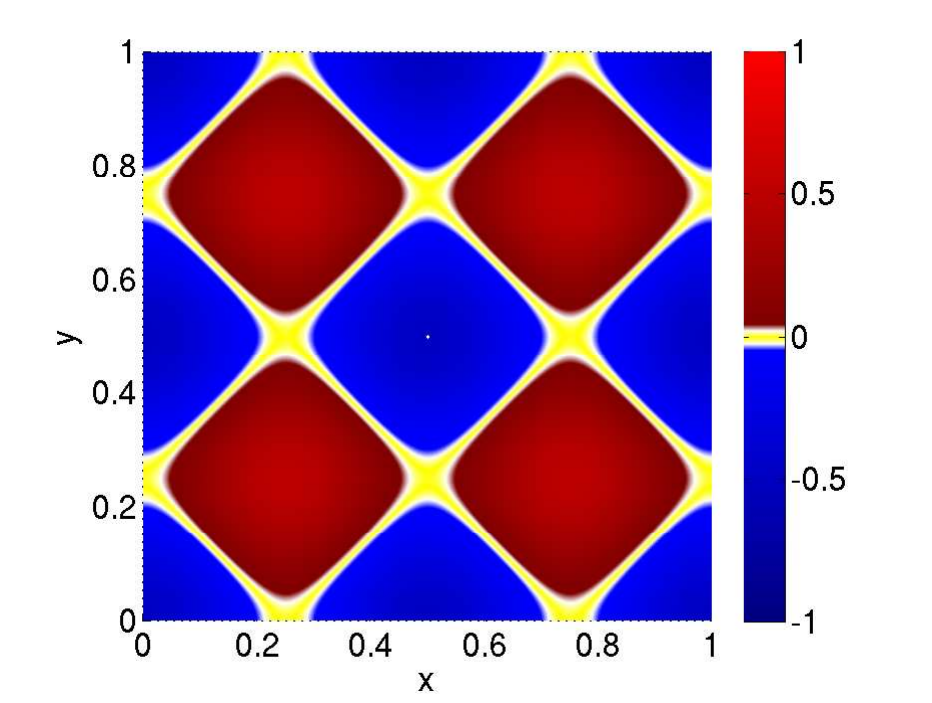}&
    \includegraphics[height=4cm]{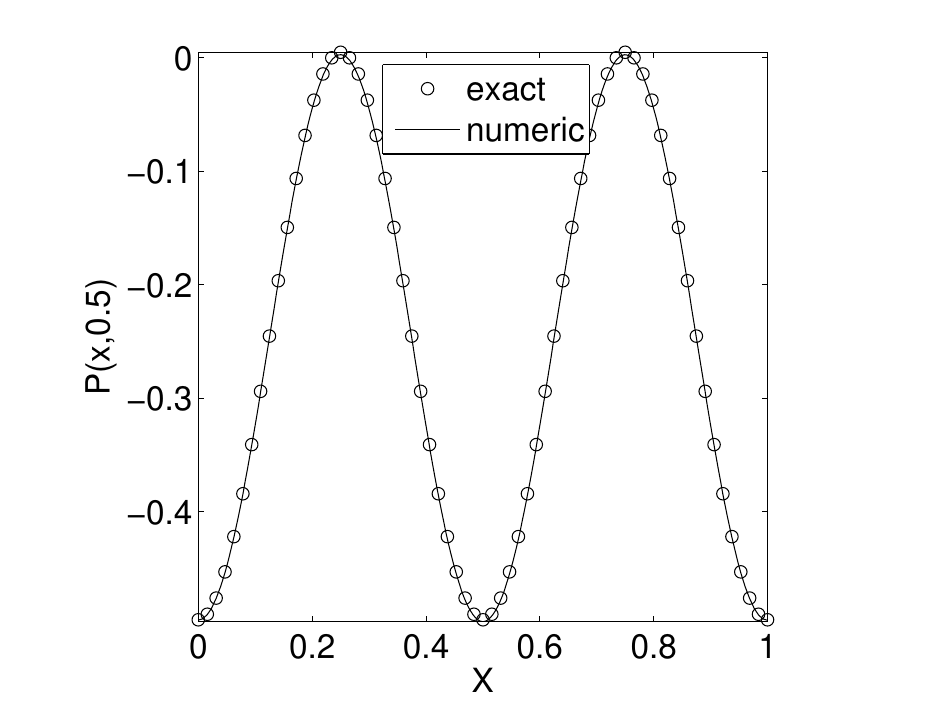}\\
    $(a)$ & $(b)$\\
    \includegraphics[height=4cm]{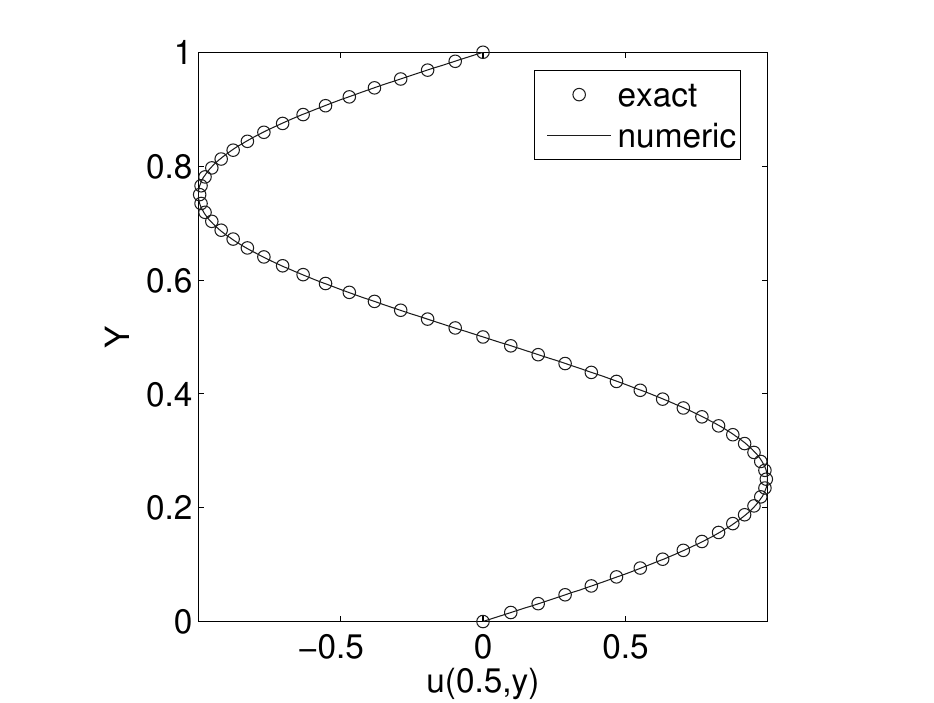}&
    \includegraphics[height=4cm]{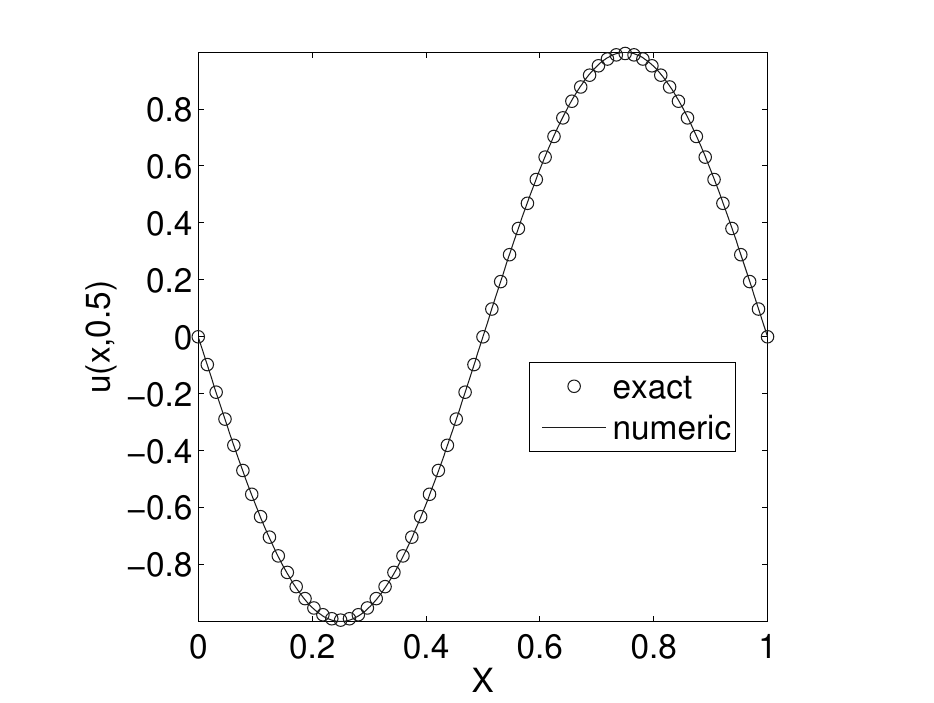}\\
    $(c)$ & $(d)$\\
     \multicolumn{2}{c}{
       \includegraphics[height=4cm]{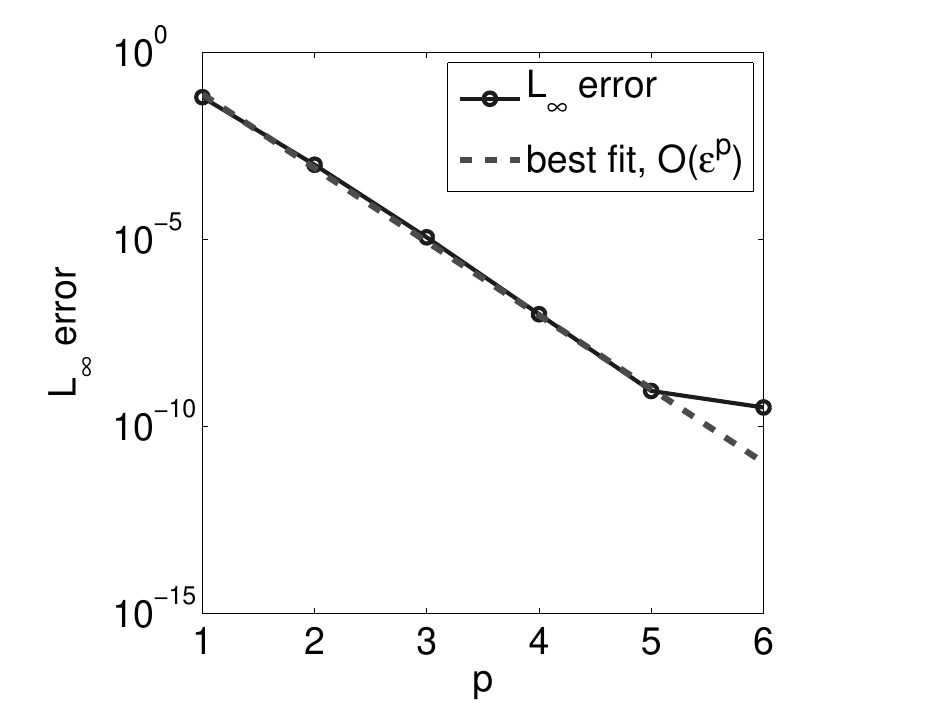}
     }\\
     \multicolumn{2}{c}{
       $(e)$
     }\\
  \end{tabular}
  \caption{Numerical solution of (\ref{eq:prj}) for $P(x,y)$, $u(x,y)$, and $v(x,y)$; $(a)$ $P(x,y)$. Computed $P(x,0.5)$, $u(0.5,y)$, and $v(x,0.5)$ are compared with the corresponding exact values in $(b)$, $(c)$, and $(d)$, respectively, where an excellent agreement is seen. \Add{$(e)$~$L_{\infty}$ error as a function of the interpolation order $p$.} }
  \label{fig:ustr}
\end{figure}

A vector field can be decomposed as the sum of a divergence free vector field and a curl free or conservative vector field, {\em i.e.} $\bm u^* = \bm u + \bm\nabla P$, where $\bm\nabla\cdot\bm u = 0$ and $P(x,y)$ is a scalar potential. 
When the incompressible Navier-Stokes equation is solved with a fractional step time integration scheme, which was originally proposed by~\citet{Chorin68}, the Helmholtz-Hodge decomposition is employed~\cite{Tannehill97}. In order to verify the development, 
let $u^*$ be a given velocity field such that $\bm\nabla\cdot\bm u^*\ne 0$, we have
\begin{equation}
  \label{eq:prj}
  \bm\nabla^2 P = \bm\nabla\cdot\bm u^*,\quad\bm\nabla P\cdot\hat n = 0,
\end{equation}
where $\hat n$ represents the outward unit vector on a corresponding boundary.
Eq.(\ref{eq:prj}) is one application of the Poisson model~(\ref{eq:bvp}) in Fluid Dynamics, where $\bm\nabla\cdot\bm u^*$ appears as if a charge distribution. Eq.~(\ref{eq:prj}) is solved to  compute the divergence-free component according to 
$$\bm u = \bm u^*-\bm\nabla P.$$
In order to verify the accuracy of the numerical solution, consider the following manufactured velocities
$$ u^*=\overbrace{-\cos(2\pi x)\sin(2\pi y)}^{\hbox{Taylor-Green velocity}}+\underbrace{\pi\sin(4\pi x)}_{\hbox{noise}},\quad v^*=\overbrace{\sin(2\pi x)\cos(2\pi y)}^{\hbox{Taylor-Green velocity}}+\underbrace{\pi\sin(4\pi y)}_{\hbox{noise}},$$ 
which are constructed by adding noise terms into the the Taylor-Green vortex solution of the incompressible Navier-Stokes equation -- a commonly used CFD toy model~({\em e.g.}~\cite{Jobelin2006}).  We can verify that
$$ P = -\frac{1}{4}\left[\cos(4\pi x)+\cos(4\pi y)\right]$$
is a solution of~(\ref{eq:prj}) in $[0,1]\times[0,1]$,
and
$$ u=-\cos(2\pi x)\sin(2\pi y),\quad v=\sin(2\pi x)\cos(2\pi y).$$
Eq~(\ref{eq:prj}) is discretized with the proposed method, and the resulting system of equations is solved with a GMRES method~\cite{Brown94} using a tolerance $10^{-4}$ on the relative residual error. The computed potential $P^{\mathcal N-1}(x,y)$ with $\mathcal N = 65\times 65$ is presented in Fig~\ref{fig:ustr}($a$), which is compared with the exact solution in Fig~\ref{fig:ustr}$(b)$. The numerical solution $u^{\mathcal N-1}(0.5,y)$ is compared with the exact solution $u(0.5,y)$ in Fig~\ref{fig:ustr}$(c)$ as well as $v^{\mathcal N-1}(0.5,y)$ is compared with the exact solution $v(0.5,y)$ in Fig~\ref{fig:ustr}$(d)$. 

From these graphical illustrations in Fig~\ref{fig:ustr}, it is hard to see the difference between the exact and the numerical solution with a naked eye. \Add{To show a quantitative assessment, the maximum error, $|P(x,y)-P^{\mathcal N}(x,y)|_{\infty}$ has been computed for $p=1\ldots 6$. As depicted in Fig~\ref{fig:ustr}$(e)$, the error agrees with the theoretical estimate $\mathcal O(\varepsilon^p)$ (the jump at $p=6$ is an accumulation of the round-off error).} These numerical experiments verify the performance of the present development for solving a Poisson equation with Neumann boundary conditions. 
\subsection{Poisson equation in complex geometry}
The DD subdivision is not restricted on the real line or to a regularly sampled data~\cite{Dubuc89}. The original development considers a function sampled on integers, which is extended iteratively to a continuous functions.  \citet{Sweldens95} (and similar works) studied the iterative interpolation on irregular meshes and complex geometries. In the present implementation, the domain can be a collection of rectangles (or rectangular prism in 3D), such as a domain with a rectangular hole. 

\begin{figure}
  \centering
  \begin{tabular}{cc}
    \includegraphics[height=4cm]{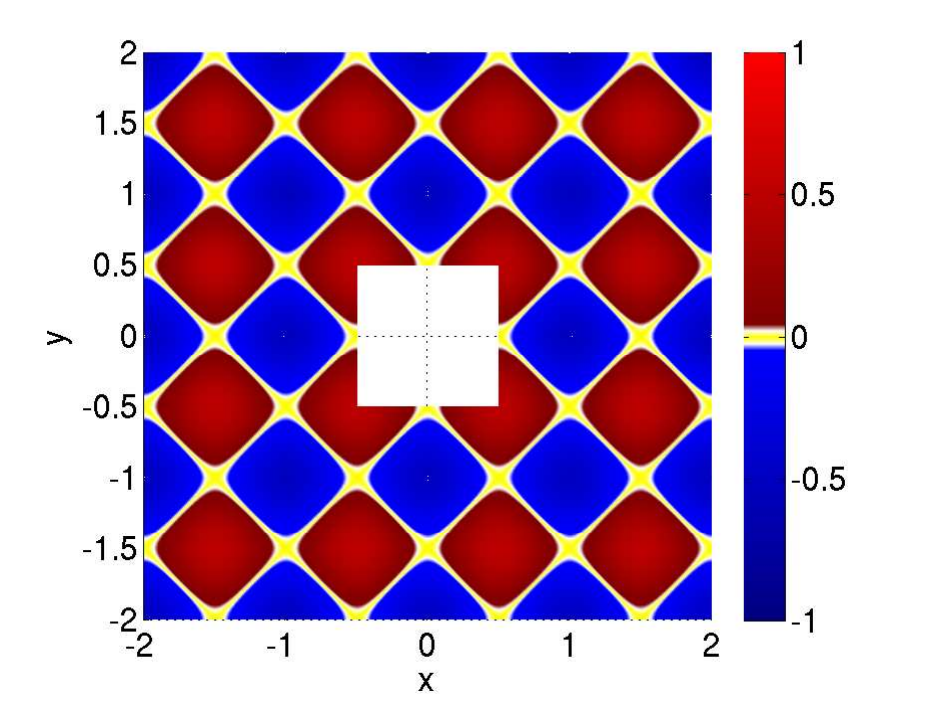}&
    \includegraphics[height=4cm]{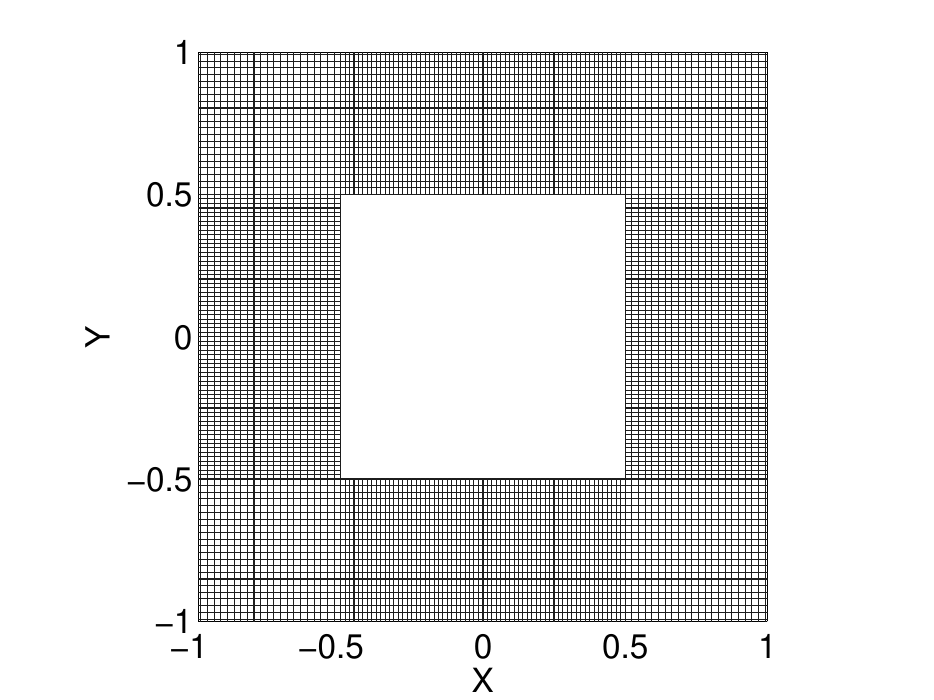}\\
    $(a)$ & $(b)$\\
  \end{tabular}
  \caption{Numerical solution of~(\ref{eq:prj}) in a domain with a hole using a nonuniform mesh. $(a)$ Solution, $(b)$ mesh; only a portion of the mesh is shown for clarity.}
  \label{fig:sqr}
\end{figure}

To demonstrate the present collocation method on a domain with a hole, we have solved~(\ref{eq:prj}) in the domain $[-2,2]\times[-2,2]\backslash[-0.5,0.5]\times[-0.5,0.5]$, which has a hole. Since the exact solution is known, we can assess the accuracy. Without going to further details, let us present the numerical solution and the associated mesh. For a better visualization, we have presented only a portion of the mesh, where the solution has been presented in the entire domain in Fig~\ref{fig:sqr}. 

Note that the numerical simulation of a fluid flow in a complex geometry is itself a challenging and independent research topic. In this work, we only want to present primary results on the potential benefits of the scaling function based collocation approach toward this direction. We are more interested to study performance of the present collocation method for simulating some representative CFD simulations.

%
\subsection{A dynamical core for simulating two-dimensional flows}\label{sec:brg}
%
This section extends the proposed discretization methodology to solve the following nonlinear advection-diffusion model:
\begin{equation}
  \label{eq:bgr}
  \frac{\partial\bm u}{\partial t} + \bm u\cdot\bm\nabla\bm u = \nu\nabla^2\bm u,
\end{equation}
where $\bm u=[u, v]^T$, which works as a dynamical core. A numerical treatment of the advection term, $\bm u\cdot\bm\nabla\bm u$, remains challenging. If $\nu\ll 1$, explicit schemes usually require a shock capturing method, such as an upwind  or an essentially non-oscillatory (ENO) finite/difference volume scheme~\cite{Harten97,Tannehill97}.  Such a scheme introduces `artificial numerical dissipation', where the time step ($\Delta t$) is also restricted through the CFL (Courant~Friedrichs~Lewy) stability criterion, $\Delta t < \frac{||\Delta \bm x||}{||\bm u||}$. Since the present method discretizes $\bm u\cdot\bm\nabla\bm u$ through eq~(\ref{eq:trld1}), it requires further technical development in order to implement upwind/ENO schemes. In contrast, an implicit treatment iteratively resolves the simultaneous nonlinear dependence between each component of $\bm u$.

Many authors consider~(\ref{eq:bgr}) as a proof of concept for their numerical scheme~\cite{Fletcher83,Kannan2012,Liao2010,Xie2010,Zhu2010}. This nonlinear advection diffusion equation is often called 2D Burger's equation because of its similarity with the Burger's equation. \citet{Kannan2012} and \citet{Liao2010} used the Hopf--Cole transformation to eliminate the nonlinear term, and as a result, solved a diffusion equation in order to obtain the solution of~(\ref{eq:bgr}). \citet{Xie2010} studied a compact finite difference scheme for Burger's equation. \citet{Zhu2010} examined the adomain decomposition method for solving eq~(\ref{eq:bgr}). Thus, a number of reference studies are available, showing that  on the avenue of solving~(\ref{eq:bgr}), the search for the best method of solving~(\ref{eq:bgr}) remains active.

A Poisson like nonlinear system of equations, $\mathcal L(\bm u^{n+1}) = \bm f$, such that
$$
\underbrace{-\nu\nabla^2\bm u^{n+1} + \bm u^{n+1}\cdot\bm\nabla\bm u^{n+1} + \frac{2}{\Delta t}\bm u^{n+1}}_{\mathcal L(\bm u)}
=
\underbrace{\nu\nabla^2\bm u^{n} - \bm u^{n}\cdot\bm\nabla\bm u^{n} + \frac{2}{\Delta t}\bm u^{n}}_{\bm f}
$$
has been obtained by discretizing~(\ref{eq:bgr}) in time with a second order Crank-Nicolson~(CN) method. To compute $\bm u^{n+1}$ at each time step by solving the nonlinear system, $\mathcal L(\bm u^{n+1}) = \bm f$, a trial solution of the form~(\ref{eq:trl}) is assumed for each component of $\bm u^{n+1}$. Evaluating the inner product $\langle\mathcal L(\bm u^{n+1}) - f,\tilde\varphi(\bm x-\bm x_k\rangle=0$, a simultaneous nonlinear system $\mathcal L(\bm u^{n+1}(\bm x_k)) = f(\bm x_k)$ of $2\mathcal N$ algebraic equations is obtained, which may be stated by $\mathcal L(\bm u) = f$, for simplicity, including appropriate boundary conditions. The discrete nonlinear system $\mathcal L(\bm u) = \bm f$ has been solved with the Newton's method (see~\cite{Ortega70}), 
$$
\bm u^{j+1} = \bm u^j + \bm s^j\quad\hbox{such that}\quad
\mathcal J(\bm u^j)\bm s^j = f-\mathcal L(\bm u^j),
$$
where one needs the Jacobian matrix $\mathcal J(u^j)$ at each iteration $j$. 
Clearly, on a mesh of $\mathcal N$ nodes, the computation of the product $\mathcal J(\bm u^j)\bm s^j$ between the Jacobian matrix $\mathcal J(\bm u^j)$ and the error vector $\bm s^j$ requires $\mathcal O(\mathcal N^2)$ operations. Therefore, the implicit treatment of the advection term is too expansive for large scale CFD applications. In order to bring the computational complexity to $\mathcal O(\mathcal N)$, we have considered the Fr\'{e}chet derivative,
$$
\lim_{\eta\bm s^j\rightarrow 0}\frac{||\mathcal L(\bm u^j + \eta\bm s^j)-\mathcal L(\bm u^j) - \mathcal J||}{||\eta\bm s^j||}
$$
which leads to
$$
\mathcal J\bm s^j \approx \frac{\mathcal L(\bm u^j + \eta\bm s^j) - \mathcal L(\bm u^j)}{\eta},
$$
and as a result, $\mathcal J\bm s^j$ can be approximated with $\mathcal O(\mathcal N)$ operations, where $\eta$ is a small number. Numerical experiments suggested that $\eta\le 10^{-4}$ is sufficient for this example. Although \citet{Knoll2004} have been reviewed this approach of solving nonlinear system of equations for multiphysics problems, in the area of CFD, it is not a commonly adopted technique (see also, \cite{Alam2012}). We have considered several numerical experiments to study the convergence of the solution. A representative example has been presented below. \Add{For all examples, $\bm s^j$ is computed with the GMRES method, along with a Jacobi preconditioning.} 

Using the same initial and boundary conditions as the reference solution ({\em e.g.} problem $1$) presented by~\citet{Zhu2010}, eq.~(\ref{eq:bgr}) has been solved in the domain $[0,1]\times [0,1]$, where the exact solutions are given by
$$
u = \frac{3}{4}-\frac{1}{4(1+\exp(\nu(-t-4x+4y)/32)},\quad
v = \frac{3}{4}+\frac{1}{4(1+\exp(\nu(-t-4x+4y)/32)}.
$$
We have analyzed this example with a time step, $\Delta t$, between $10^{-1}$ and $10^{-4}$, where for each $\Delta t$, the resolution varies between $33\times 33$ and $129\times 129$. Thus, we have, $12.8\times 10^{-3}\le\hbox{ CFL }\le 12.8$. With CFL~$=12.8$, $\Delta t=10^{-1}$, and $\nu=1.25\times 10^{-2}$, the maximum absolute error $5.05\times 10^{-4}$. In comparison, \citet{Zhu2010} reported a maximum absolute error $7.5\times 10^{-4}$ with $\Delta t=10^{-4}$ and  $\nu=1.25\times 10^{-2}$. This comparison with a $\Delta t$ that is $10^3$ times larger than what was used by~\citet{Zhu2010}, indicates that the present multiresolution collocation method is able to refine the mesh at CFL~$=12.8$ without reducing the time step, and at this high CFL, the error bound is equivalent to that of the scheme of~\citet{Zhu2010}. This explains the performance of the present method for the nonlinear advection-diffusion problem, and the result should be considered carefully because it does not claim that the present method is superior to that of~\cite{Zhu2010}.

\begin{figure}
  \centering
  \begin{tabular}{cc}
    \includegraphics[height=4cm]{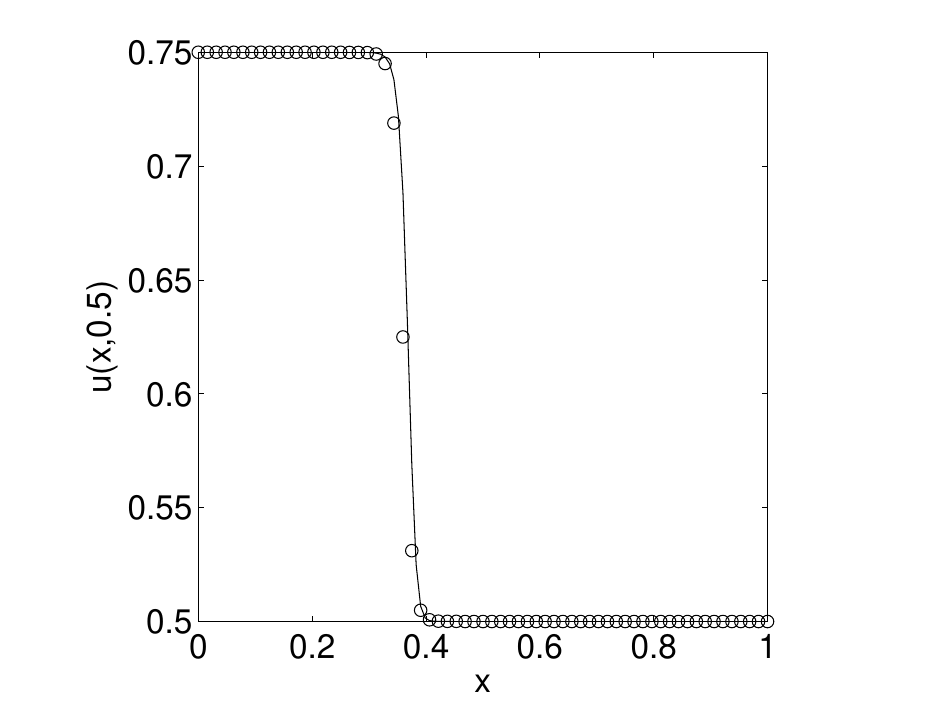}&
    \includegraphics[height=4cm]{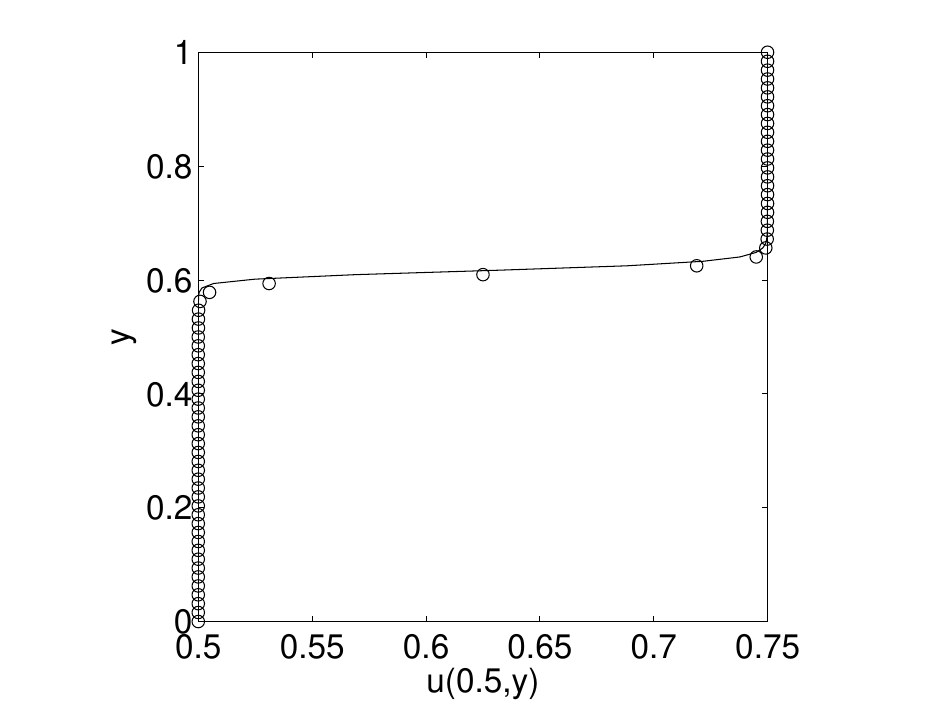}\\
    \includegraphics[height=4cm]{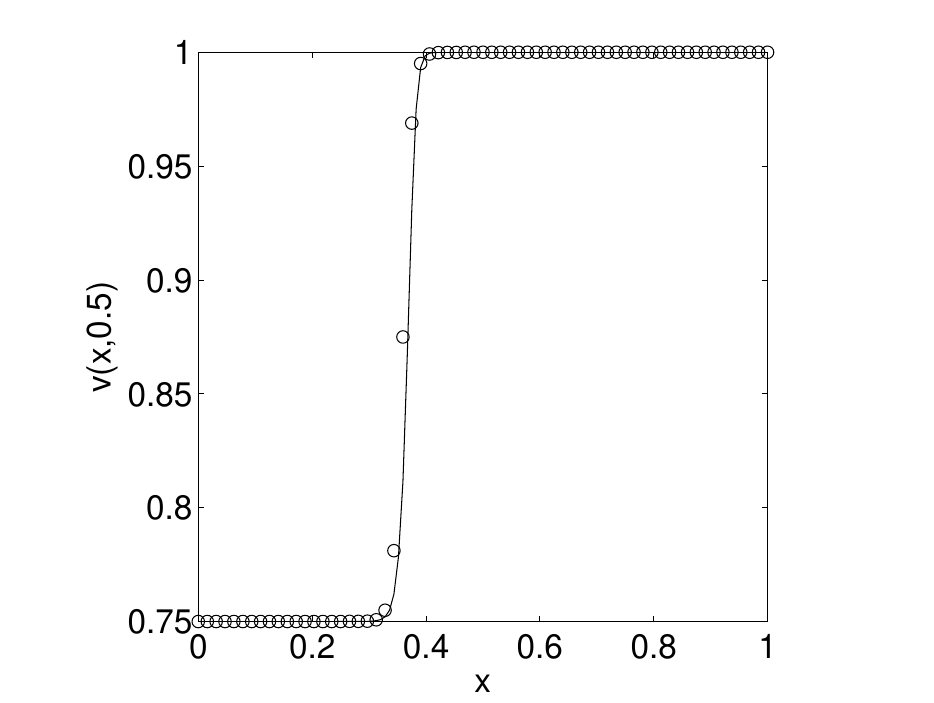}&
    \includegraphics[height=4cm]{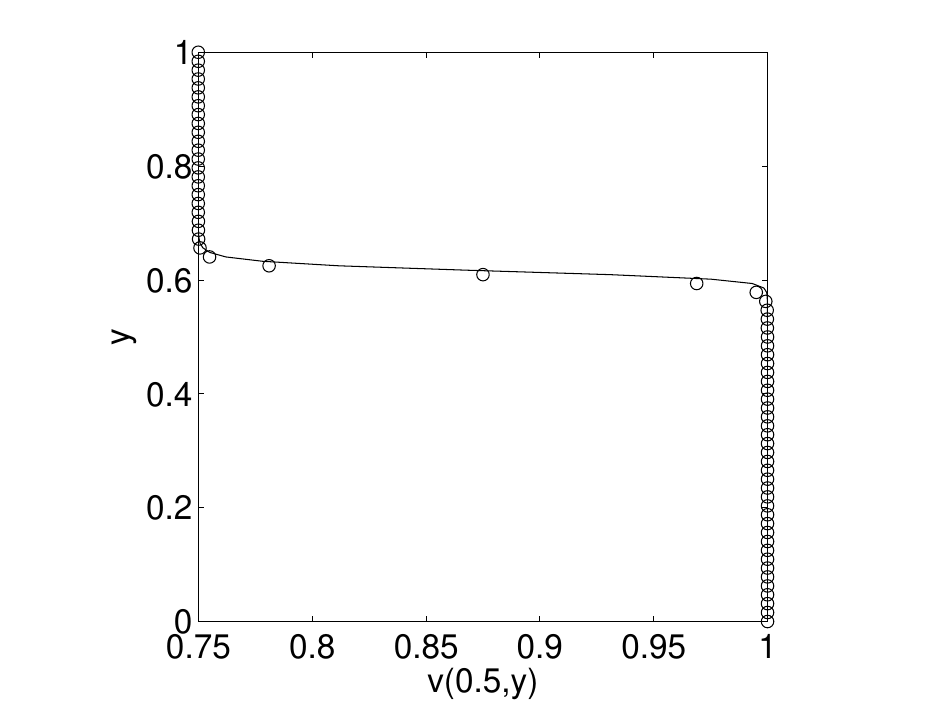}\\
  \end{tabular}
  \caption{Numerical solutions of~(\ref{eq:bgr}) $(a)$ $u(x,0.5)$, $(b)$ $u(0.5,y)$, $(c)$ $v(x,0.5)$, and $v(0.5,y)$.}
  \label{fig:bgr}
\end{figure}

In Fig~\ref{fig:bgr}, we compare numerical solution with the exact solution~(e.g.,~\cite{Zhu2010}). The plots include $u(x,0.5)$, $u(0.5,y)$, $v(x,0.5)$, and $v(0.5,y)$. The excellent agreement between the exact and the numerical solutions with no visible oscillation encourages the methodology to the field of Computational Fluid Dynamics. 

\subsection{Verification with a two-dimensional shear driven flow}
A shear-driven flow is a circulation in a confined box~\cite{Ghia82,Perrin2006,Ghadimi2013}, where an imposed shear stress drives the fluid, and is a classical test problem for the assessment of CFD codes. The simulation of a shear driven flow using the incompressible Navier-Stokes equation is a challenging endeavor in the field of Computational Fluid Dynamics. \citet{Weinan95} discussed the occurrence of artificial numerical boundary layer if a classical fractional step method is employed to solve the incompressible Navier-Stokes equation (see also, \cite{San2013}). In this article, we do not have enough room to address these unresolved challenges with this fractional step projection method; however, we aim to demonstrate the potential of the present discretization technique to the field of CFD, using a simulation of the classical shear driven flow. This is an incompressible flow in a square cavity $[0,1]\times[0,1]$ with no slip conditions $u=0=v$ on boundaries, $x=0$, $x=1$, and $y=0$. To model the imposed shear stress, Dirichlet type boundary conditions, $u=1$ and $v=0$, have been used on the boundary, $y=1$. Since we cannot derive the exact solution for the shear driven flow, a reference simulation may be used to verify our simulation. \citet{Ghia82} and \citet{Ghadimi2013}) examined a similar shear driven flow, using the steady-state vorticity equation (see eqs(1-2) of~\cite{Ghadimi2013}). Although the present numerical method and the set of equations are different than those used by~\citet{Ghia82} and~\citet{Ghadimi2013}, these reference results are useful feedback for assessing a simulation of shear driven flow using the proposed collocation method.

A shear driven flow can be modelled by the incompressible Navier-Stokes equation
\begin{equation}
  \label{eq:nse}
  \frac{\partial\bm u}{\partial t} + \bm u\cdot\bm\nabla\bm u = -\bm\nabla P+\frac{1}{Re}\nabla^2\bm u,\quad\bm\nabla\cdot\bm u = 0,
\end{equation}
where $\mathcal Re = \frac{\rho_0UL}{\mu}$, $\rho_0$ is a reference density, $\mu$ is the dynamic viscosity, $U$ is a velocity scale, $L$ is a length scales, and the dimensional density, $\tilde\rho$, has been perturbed by $\tilde\rho=\rho_0+\rho$ such that $|\frac{\rho}{\rho_0}|\ll 1$, which is the same as the Boussinesq approximation. After substituting this density perturbation into the conservation of mass, $\frac{\partial\tilde\rho}{\partial t}+\bm\nabla\cdot(\bm u\tilde\rho)=0$, in order to satisfy the incompressibility condition, $\bm\nabla\cdot\bm u = 0$, we must have
\begin{equation}
  \label{eq:dst}
  \frac{\partial\rho}{\partial t} + \bm u\cdot\bm\nabla\rho + \rho\bm\nabla\cdot\bm u = 0  
\end{equation}
for the perturbation density. Following \citet{Perrin2006} and \citet{Chorin67}, the dimensionless pressure in~(\ref{eq:nse}) is given by the dimensionless equation of state, $P=\frac{\rho}{\rho_0M^2}$, where $M$ is the Mach number (see also, Chapter 11, \cite{Kundu} and Chapter~9, \cite{Tannehill97}), and we have used $\rho_0U^2$ and $\rho_0M^2$, as a scale for pressure and density, respectively. 

The present model~(\ref{eq:nse}-\ref{eq:dst}) of the incompressible flow is an equivalent extension to the artificial compressibility method that was proposed by~\citet{Chorin67}. However, in the method of~\citet{Chorin67}, an artificial time derivative was added; {\em i.e.} $\frac{\partial\rho^*}{\partial t^*}+\bm\nabla\cdot\bm u=0$ (see also, eq~9.135 of~\cite{Tannehill97} and \cite{Nithiarasu2003}), and thus, a steady state $\frac{\partial\rho^*}{\partial t^*}=0$ is needed at each physical time step to update $P=\beta\rho^*$, where $\beta$ is a model parameter. \citet{Perrin2006} solved eq~(\ref{eq:dst}) along with the compressible Navier-Stokes equation using the explicit MacCormack finite difference method, which is a conditionally stable scheme, and showed an excellent result with the shear driven flow simulation. In the present work, we have extended this approach in such a way that the simultaneous dependence between the velocities, $u$ and $v$, and density perturbation $\rho$ has been resolved through an iterative algorithm. 

After a temporal discretization with the Crank-Nicolson scheme, eqs~(\ref{eq:nse}-\ref{eq:dst}) leads to a nonlinear system of `Poisson like' PDEs, $\mathcal L(\bm u^{n+1}) = \bm f$, where we have defined
$$
\mathcal L(\bm u^{n+1}) = \left[
  \begin{array}{ll}
    -\frac{1}{\mathcal Re}\nabla^2\bm u^{n+1} + \bm u^{n+1}\cdot\bm\nabla\bm u^{n+1} 
    +\bm\nabla P^{n+1} + \frac{2\bm u^{n+1}}{\Delta t}\\
    \bm u^{n+1}\cdot\bm\nabla\rho^{n+1} + \rho^{n+1}\bm\nabla\cdot\bm u^{n+1} + \frac{2\rho^{n+1}}{\Delta t}
  \end{array}
\right],
$$
and
$$
\bm f = \left[
  \begin{array}{ll}
    \frac{1}{\mathcal Re}\nabla^2\bm u^{n} - \bm u^{n}\cdot\bm\nabla\bm u^{n}
    - \frac{2\bm u^{n+1}}{\Delta t}
    + \frac{2\bm u^{n}}{\Delta t}\\
    -\bm u^{n}\cdot\bm\nabla\rho^{n} - \rho^{n}\bm\nabla\cdot\bm u^{n} + \frac{2\rho^{n}}{\Delta t},    
  \end{array}
\right].
$$
As described in section~\ref{sec:brg}, a trial solution of the form~(\ref{eq:trl}) has been considered for discretizing each of $u^{n+1}$, $v^{n+1}$ and $\rho^{n+1}$ in space, and the resulting system of algebraic equation can be denoted by $\mathcal L(u) = \bm f$. This nonlinear system of $3\mathcal N$ equations has been solved at each time step with the Jacobian-free Newton-Krylov algorithm presented in section~\ref{sec:brg} (see, \cite{Knoll2004,Alam2011}). 


For this simulation, $256$ uniform rectangles in both the $x$ and $y$ directions have been used at the highest resolution, where the total number of degrees of freedom is $3\mathcal N$ with $\mathcal N = 257\times 257$. With this $\mathcal N$, we have $\Delta x=\Delta y\sim 4\times 10^{-3}$. 

A number of simulations with $\mathcal N$ between $33\times 33$ and $257\times 257$ and Reynolds number, $\mathcal Re$, between $100$ and $1\,000$ have been considered. For $\mathcal N = 257\times 257$, we have tested time steps ($\Delta t$) between $10^{-5}$ and $10^{-2}$. Analysis of the simulated data shows that the average CPU times in the wall-clock unit, for CFL numbers $0.25$ ($\Delta t=10^{-3}$) and $2.5$ ($\Delta t=10^{-2}$), are about $12$~days and $1$~day, respectively, for the same dimensionless integration time. As expected, the speed up for the implicit scheme is approximately linear with respect to time steps. Most classical CFD codes would use a CFL~$<1$ because the advection term is typically treated with an explicit scheme. The method of~\citet{Perrin2006} is further restrictive because the viscous term has also been treated explicitly. 

\begin{figure}
  \centering
    \begin{tabular}{cc}
    \includegraphics[height=4cm]{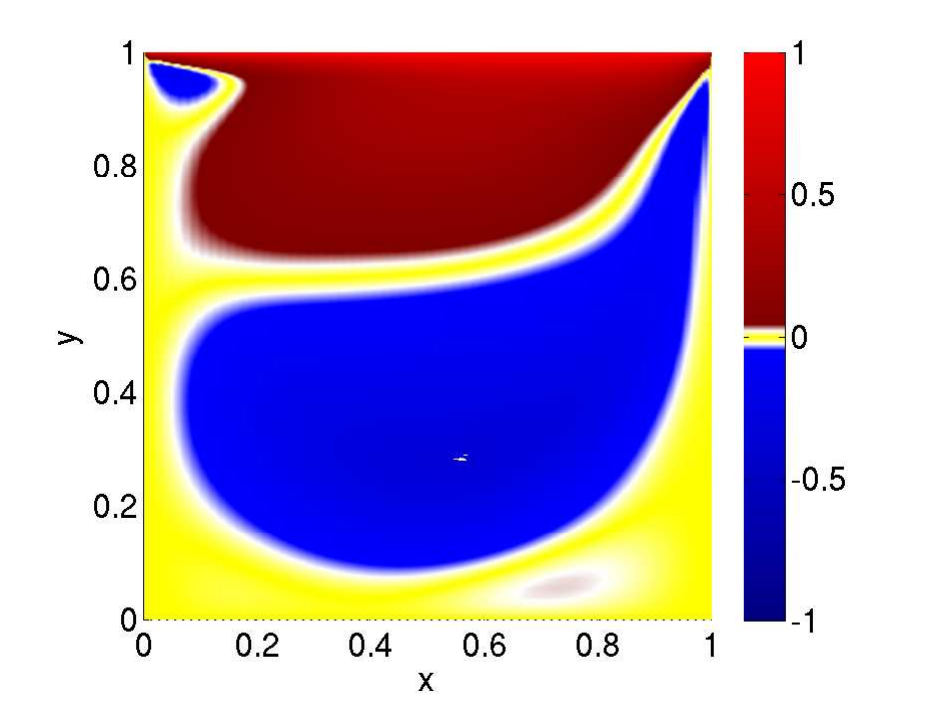}&
    \includegraphics[height=4cm]{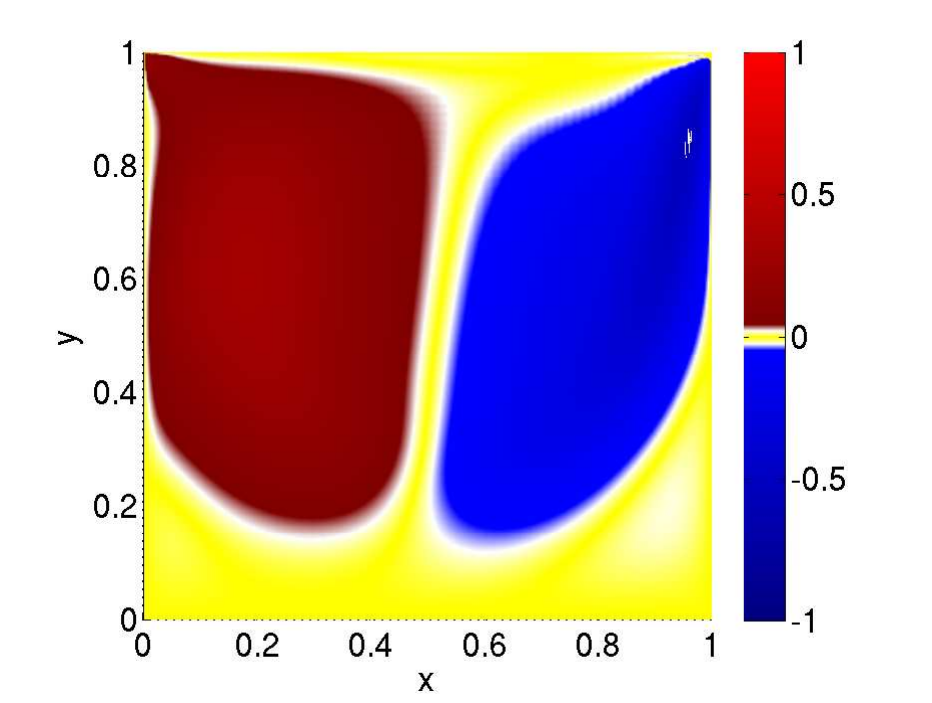}\\
    $(a)$ & $(b)$\\
     \multicolumn{2}{c}{
       \includegraphics[height=4cm]{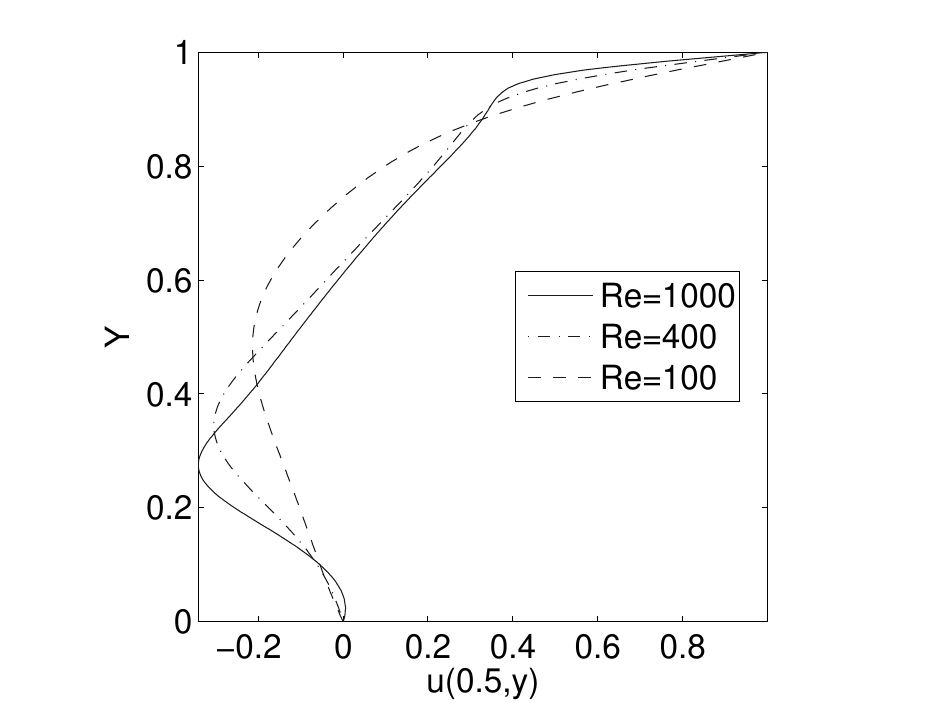}
     }\\
     \multicolumn{2}{c}{
       $(c)$
     }\\
  \end{tabular}
  \caption{Velocities for the shear driven flow. $(a)$ $u(x,y)$, $(b)$ $v(x,y)$, and $(c)$ $u(0.5,y)$ for $\mathcal Re=100$, $400$, and $1\,000$. The overall circulation in ($a$-$b$) and the velocity profiled in $c$ are in good agreement with the reference results.}
  \label{fig:shear}
\end{figure}

For $\mathcal Re=1\,000$, the simulated velocity fields are presented in Fig~\ref{fig:shear}$(a$-$b)$, showing an overall pattern of the circulation, which are in good agreement with previously reported results~\cite{Ghia82,Alam2012,Ghadimi2013}. In Fig~\ref{fig:shear}$(c)$, we have compared the simulated velocity, $u(0.5,y)$, for $\mathcal Re=100$, $400$, and $1\,000$. The pattern of the velocity profile with increasing $\mathcal Re$ is similar to what was presented by~\citet{Perrin2006}, \citet{Ghia82} and~\citet{Ghadimi2013}. Table~\ref{tab:shear} confirms a quantitative assessment of the present simulation with respect to the reference model, where the minimum values of $u(x,y)$ and $v(x,y)$ have been reported for $\mathcal Re=100$, $400$, and $1\,000$.   
\begin{table}
  \centering
  \begin{tabular}{|c|cc|cc|cc|}
    \hline
    $\mathcal Re$   & \multicolumn{2}{c|}{$u_{\min}$}& \multicolumn{2}{c|}{$v_{\min}$} \\
    \hline
    & \citet{Ghia82} & present & \citet{Ghia82} & present  \\
    $100$  & $-0.21090$ & $-0.245147$ & $-0.24533$ & $-0.245147$ \\
    $400$  & $-0.32726$ & $-0.319652$ & $-0.44993$ & $-0.549866$\\
    $1\,000$  & $-0.38289$ & $-0.346639$ & $-0.51550$ & $-0.53349$\\
    \hline
  \end{tabular}
  \caption{Comparison of the velocities $u$ and $v$ for the shear driven flow simulation. The agreement between two simulations is excellent, albeit different methods and equations have been used to  model the shear driven flow. }
  \label{tab:shear}
\end{table}

\subsection{Penetrative natural convection flow}
Atmospheric scientists utilize numerical simulations on the evolution of plumes and thermals to investigate more complicated dynamics of the atmosphere and how it influences mixing and redistribution of heat and other constituent species~\cite{Alam2011,Bryan2002,Carpenter90,Lane2008}. \citet{Carpenter90} studied a piece-wise parabolic scheme, and suggested that a more powerful multi-resolution scheme would be effective for numerical atmospheric modelling. Following the work of \citet{Lane2008}, an idealized mathematical model of penetrative convection may be written in the following primitive variable form (see, \cite{Dubois2009})
\begin{equation}
  \label{eq:bnc}
  \frac{\partial\bm u}{\partial t} + \bm u\cdot\bm\nabla\bm u
  = -\bm\nabla P+\sqrt{\frac{Pr}{Ra}}\nabla^2\bm u + 
  \left(\begin{array}{l}
      0\\1
    \end{array}
  \right)
  Ri\theta,
\end{equation}
\begin{equation}
  \label{eq:tht}
  \frac{\partial\theta}{\partial t} + \bm u\cdot\bm\nabla\theta
  = -\frac{w}{RiFr^2} + \sqrt{\frac{1}{PrRa}}\nabla^2\theta
\end{equation}
where $Pr=\nu/\kappa$ is the Prandtl number, $Ra=\beta H^3\Delta\theta g/(\nu\kappa)$ is the Rayleigh number, $Ri=\Delta\theta Hg/(\theta_0U^2)$ is the Richardson number, $Fr = U\sqrt{\theta_0}/(H\sqrt{g\frac{\partial\bar\theta}{\partial z}})$ is the Froude number, $\nu$ is the kinematic viscosity, $\kappa$ is the coefficient of heat diffusion, $\theta$ is the potential temperature, $\bar\theta(z)$ is a prescribed vertical profile of the potential temperature, $U$ is a characteristic velocity scale, $H$ is a characteristic length scale, $\theta_0$ is a reference temperature, $\Delta\theta$ is a scale for $\theta$, $\beta$ is the thermal expansion coefficient, $\frac{\partial\bar\theta}{\partial z}$ is a prescribed vertical rate of variation of $\theta$, and $g$ is acceleration due to gravity. For the present idealized mode, temperature and potential temperature are equivalent.

We have solved~(\ref{eq:dst}-\ref{eq:tht}) in a vertical cross section $[x_{\min},x_{\max}]\times[z_{\min},z_{\max}]=[-10,10]\times[0,10]$ of the volume $\Omega$, where $z$ coordinate is parallel to the direction of the gravitational force. The boundary conditions are Dirichlet type on the $z_{\min}$ boundary and Neumann type on all other boundaries. Initially, the fluid is assumed stationary, and the perturbation temperature $\theta$ is assigned to have a localized bubble, $\theta(x,z,0)=\exp(-((x-x_c)^2+(z-z_c)^2)/\nu_1)$, closed to the $z_{\min}$ boundary (see, \cite{Alam2011,Carpenter90}, for a detailed expression). For the reported simulation, we have used $Ri=0.1$, $Fr=10^2$, $Pr=0.71$, $Ra=10^5$, $\mathcal N=129\times 129$, $\Delta t=10^{-2}$.

In order to validate the simulation, the following energy balance laws have been adopted. 
The kinetic and potential energies
$$
E_k = \frac{1}{2}\int\limits_\Omega(u^2+w^2)dV,
\quad\hbox{and}\quad 
E_p = \int\limits_\Omega(z_{\max}-z)\theta dV,
$$
satisfy (see, \cite{Winters2009})
$$
\frac{dE_k}{dt} = \overbrace{\int\limits_\Omega w\theta dV}^{\hbox{production}} -\underbrace{\epsilon\sqrt{\frac{Pr}{Ra}}}_{\hbox{dissipation}},
\quad\epsilon=\int\limits_\Omega|\bm\nabla u|^2+\bm\nabla w|^2 dV
$$
and
$$
\frac{dE_p}{dt} = -\underbrace{\int\limits_\Omega w\theta dV}_{\hbox{conversion}} 
+ \overbrace{\frac{\theta_{z_{\max}}-\theta_{z_{\min}}}{z_{\max}-z_{\min}}\frac{1}{\sqrt{RaPr}}}^{\hbox{production}},
$$ 
respectively.
These energy equations quantify the rate of production of $E_p$, the conversion from $E_p$ to $E_k$, and the rate of kinetic energy dissipation, $\epsilon$, thereby making a steady state energy balance. One expects that an effective numerical simulation resolves such an energy balance (see, \cite{Alam2011,Carpenter90,Winters2009}).

\begin{figure}[hbtp]
  \centering
    \begin{tabular}{cc}
    \includegraphics[height=4cm]{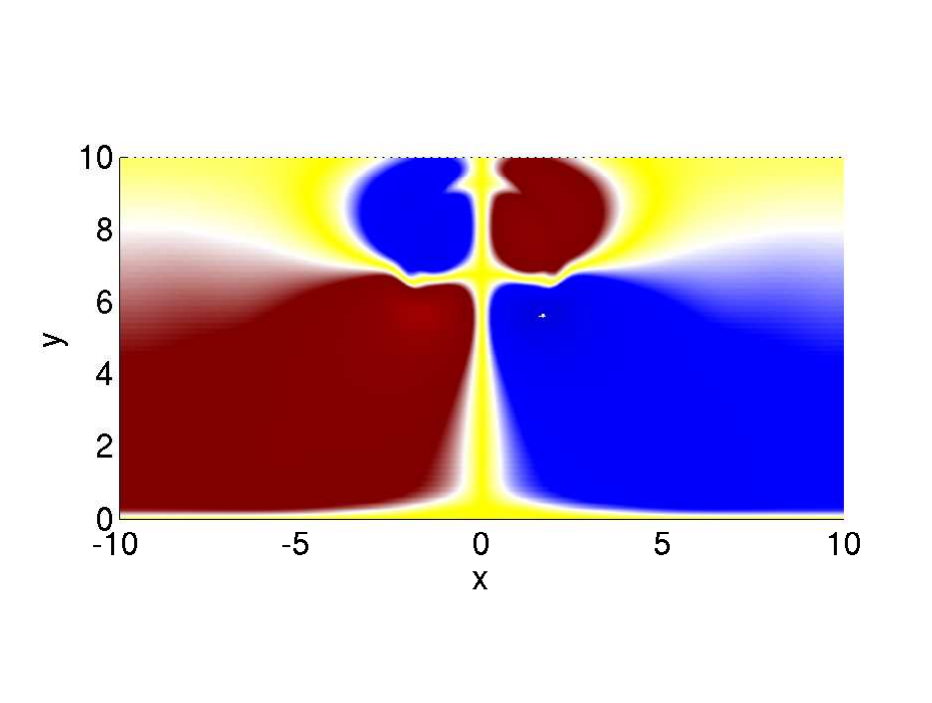}&
    \includegraphics[height=4cm]{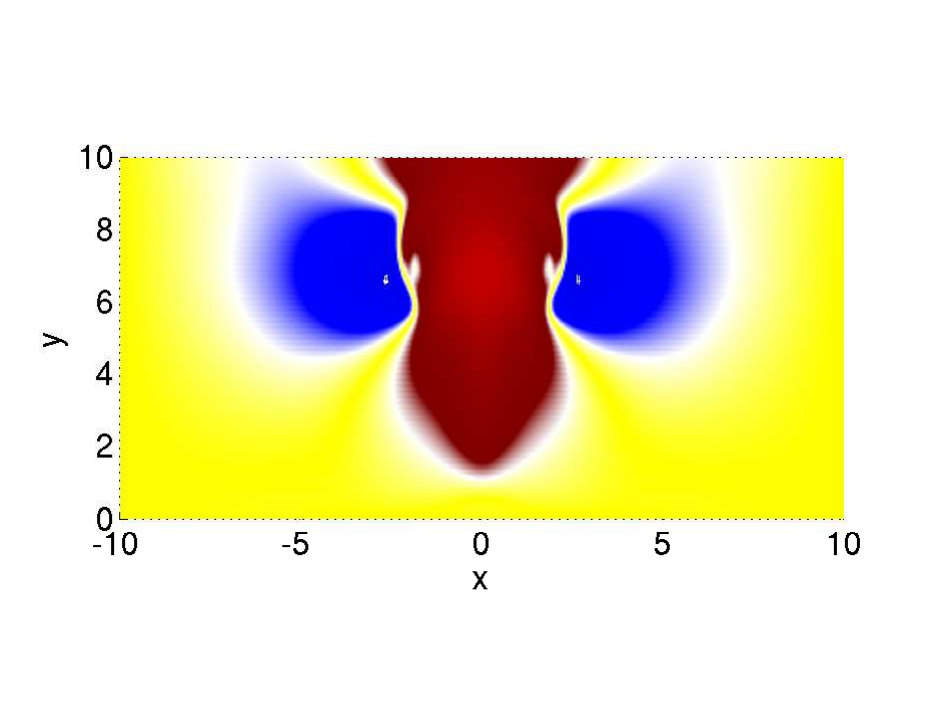}\\
    $(a)$ & $(b)$\\
     \multicolumn{2}{c}{
       \includegraphics[height=4cm]{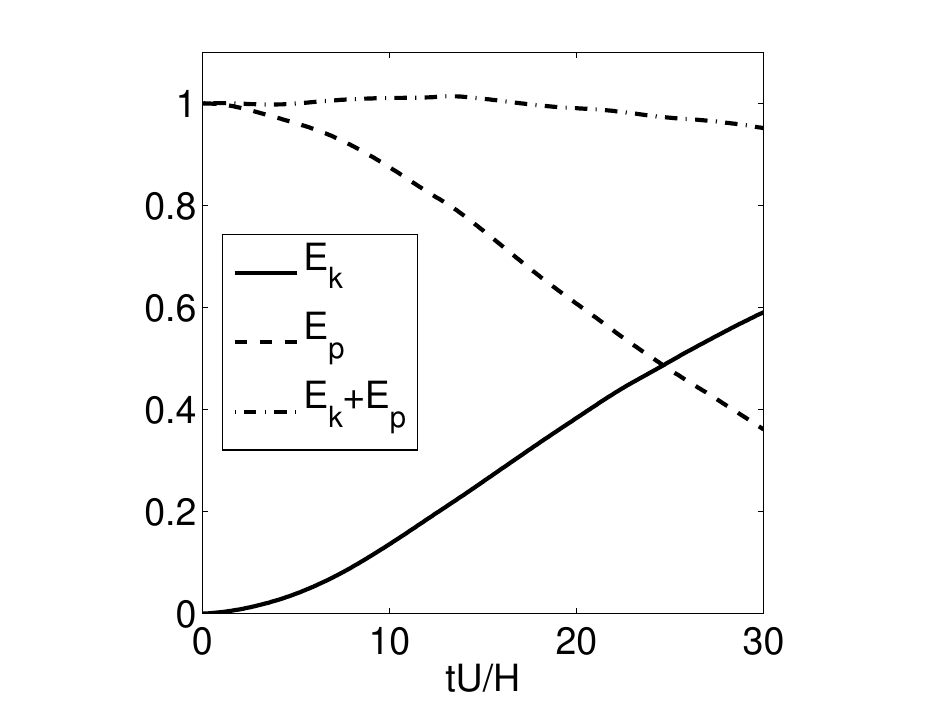}
     }\\
     \multicolumn{2}{c}{
       $(c)$
     }\\
  \end{tabular}
  \caption{Energy balance, showing that total energy is conserved, where potential energy is converted to kinetic energy}
  \label{fig:egy}
\end{figure}

In this experiment, our objective is to verify whether the present method demonstrates conservation of energy, which is one important aspect of efficient numerical approaches for atmospheric modelling. Verifying this energy conservation confirms the validity of the overall approach.  The color filled contour plots of $u(x,z,30)$ and $w(x,z,30)$ in Fig~\ref{fig:egy}($a$-$b$) do not exhibit numerical artifact, and have an excellent qualitative agreement with the velocity field reported by~\citet{Carpenter90}. However, the energy plot in Fig~\ref{fig:egy}$(c)$ provides a more quantitative measure on the accuracy of the simulation. The time evolution of $E_k$, $E_p$, and $E_k+E_p$ have been reported in Fig~\ref{fig:egy}$(c)$. Clearly, the potential energy, $E_p$, decreases in time as a result of the potential energy conversion into kinetic energy, which is seen from the increasing plot of $E_k$, as well as the total energy, $E_k+E_p$, remains approximately constant. The result on the energy conservation in Fig~\ref{fig:egy}$(c)$ has an excellent agreement with the corresponding result reported by~\citet{Carpenter90}. This result indicates the effectiveness of the present model to meteorological simulations. 

\begin{figure}[hbtp]
  \centering
    \begin{tabular}{cc}
    \includegraphics[height=5cm]{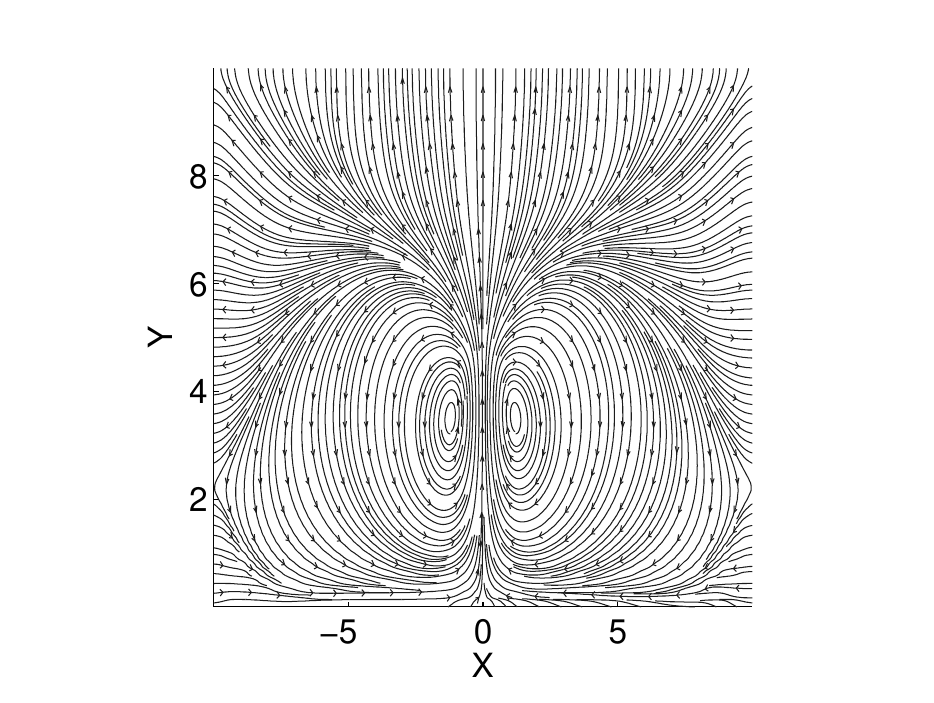}&
    \includegraphics[height=5cm]{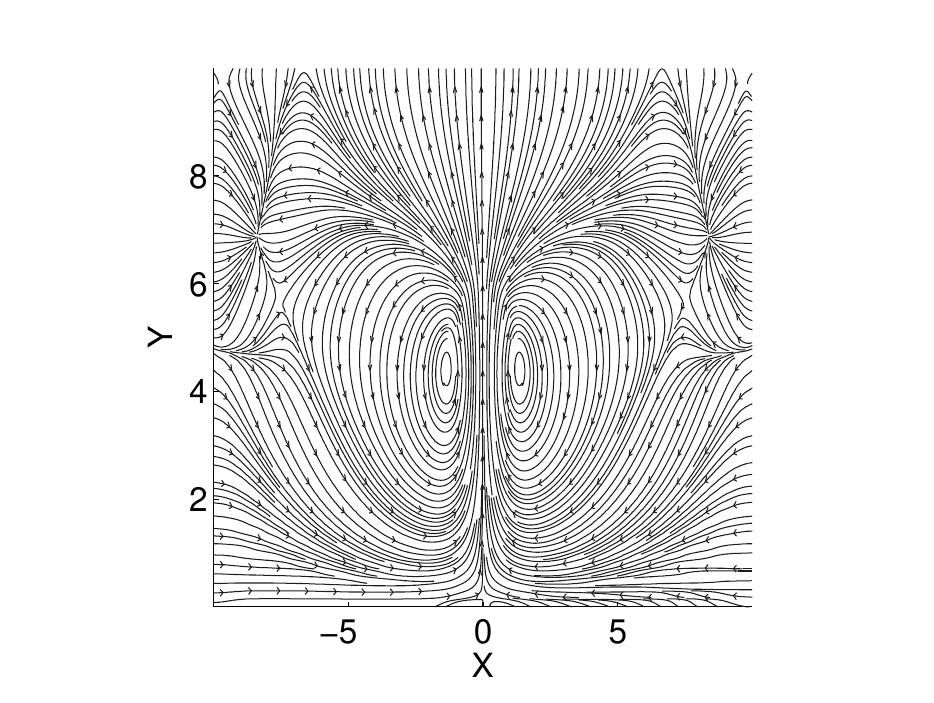}\\
    $(a)\,~t=10$ & $(b)\,~t=15$\\
    \includegraphics[height=5cm]{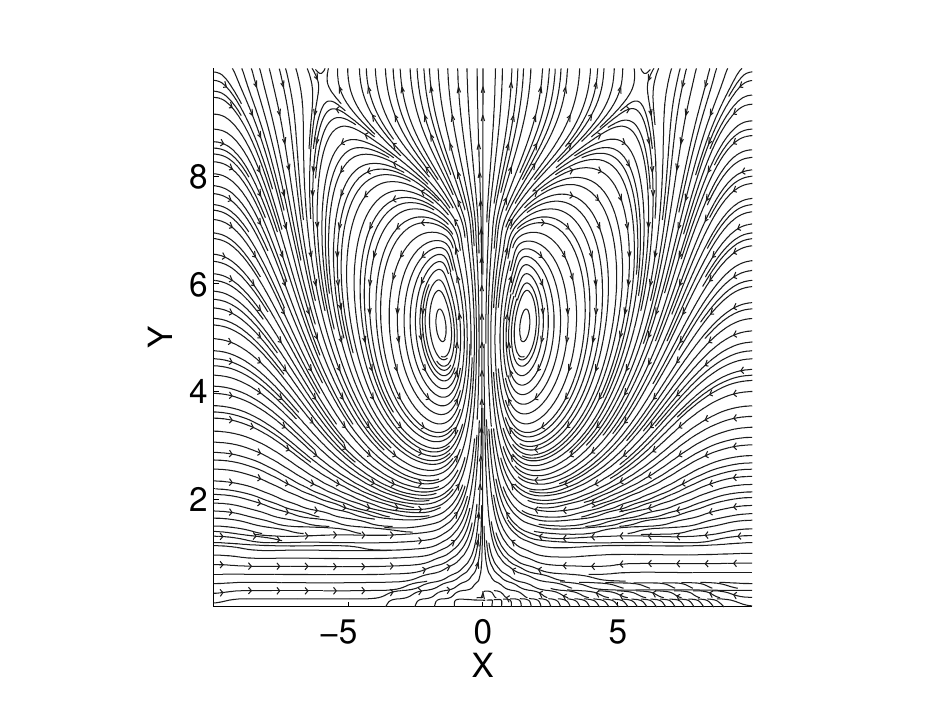}&
    \includegraphics[height=5cm]{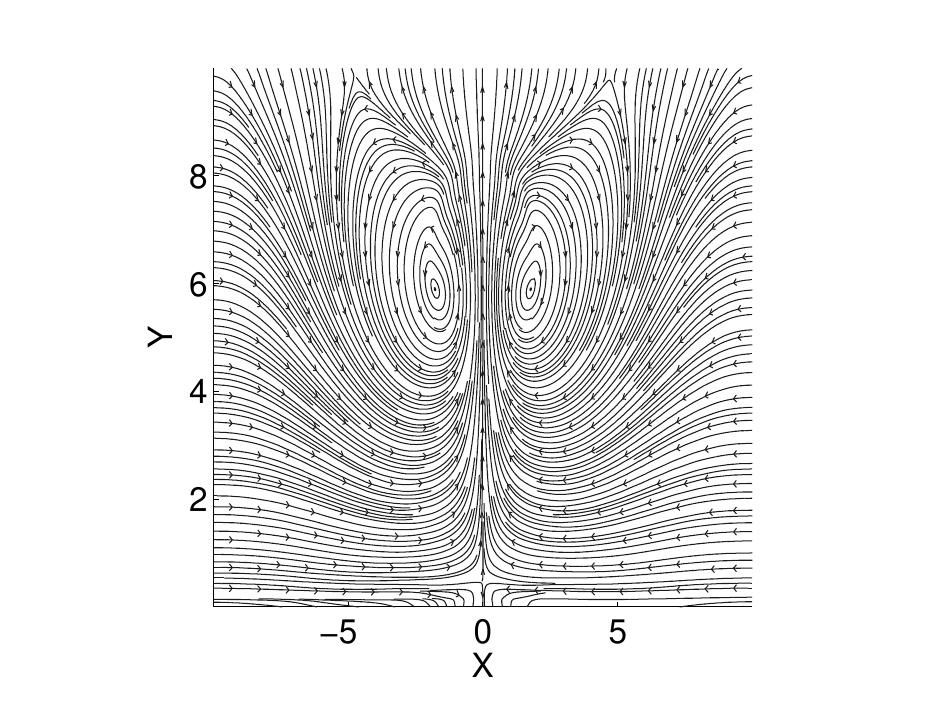}\\
    $(c)\,~t=20$ & $(d)\,~t=25$\\
     \multicolumn{2}{c}{
       \includegraphics[height=5cm]{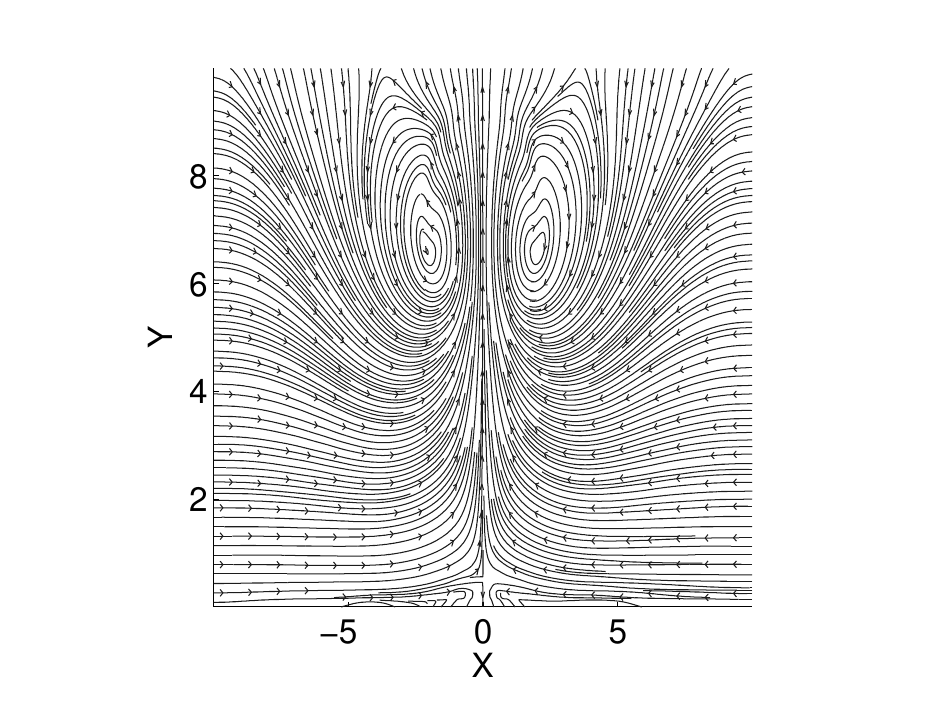}
     }\\
     \multicolumn{2}{c}{
       $(c)\,~t=30$
     }\\
  \end{tabular}
  \caption{Streamlines for the natural convection flow at increasing times, $t=10$, $t=15$, $t=20$, $t=25$, and $t=30$.}
  \label{fig:ctr}
\end{figure}

In order to provide further insight into the quality of this simulation, we present the stream lines at $t=10$, $15$, $20$, $25$, and $30$, where $t$ represents a dimensionless time. These contour plots exhibit the vertical migration of two counter rotating patterns, where the left vortex is counter clockwise, and the right vortex is clockwise. The overall pattern of the circulation in good agreement with what was presented by~\citet{Carpenter90} and \citet{Lane2008}.

\section{Conclusion and future direction}\label{sec:conc}
This article outlines a numerical simulation methodology, where partial derivatives have been discretized with a weighted residual collocation method that is based on the interpolating scaling functions, and a fully implicit time integration scheme has been studied following the artificial compressibility method~\cite{Chorin67,Nithiarasu2003,Perrin2006}. 

The basis for the weighted residual collocation method has been derived with the help of the iterative interpolation scheme proposed by~\citet{Dubuc89}, {\em i.e.} the DD subdivision scheme. An algorithm for computing the first and the second order derivatives of the interpolating scaling function has been presented. The performance of the weighted residual collocation method has been studied. The numerical verification has been presented with $3$ representative examples. $i)$ The Laplacian of a function has been approximated, verifying that the error agrees with the theoretical estimate. $ii)$~The electrostatic potential field has been computed form a given charge distribution, where a manufactured solution of the potential field is known. An excellent agreement between the exact and the numerical solution has been observed. $iii)$~The methodology has been tested for computing the Helmholtz-Hodge decomposition of a given vector field. These experiments confirm the accuracy of the collocation method. 

The present development on the collocation method has been extended for simulating two-dimensional fluid flow, which has been validated with three challenging examples; $i)$~the solution of the advection-diffusion equation exhibits no visible oscillation, and demonstrate a linear speed up because of the implicit scheme; $ii)$~the shear driven simulation has been validated with a reference simulation; and $ii)$~excellent energy conservation has been observed with the simulation of a penetrative convection flow.


These results provide potential feedback on constructing fast multiresolution  algorithms for simulating fluid flows. 
We are interested to extend the proposed discretization methodology to three-dimensional fluid flows. However, an advanced data structure, a parallel computing algorithm, and a multilevel solution methodology are needed for three-dimensional simulations. This work is currently underway.

\section*{Acknowledgments}
JMA acknowledges financial support from the National Science and Research Councill~(NSERC), Canada.

\bibliographystyle{apalike}
\bibliography{bibrefs}

\end{document}